\documentclass[journal]{IEEEtran}

\usepackage{cite}
\usepackage{amsmath,amssymb,amsfonts}
\usepackage{graphicx}
\usepackage{booktabs}
\usepackage{array}
\usepackage{tikz}
\usetikzlibrary{arrows.meta,positioning,fit,shapes.geometric}
\usepackage[colorlinks=false,hidelinks]{hyperref}

\graphicspath{{figures/}}

\newcommand{\Phalf}{\Phi_{1/2}}

\begin{document}

\title{Diffused-Beam Laser-Diode LiFi Under Realizable Receiver, Noise, and
Safety Constraints:\\ Design-Space Analysis and an Open Cross-Verified
Simulation Framework}

\author{Hussain~Ahmad,
        Saleem~Aslam,
        Ammara~Nasim,
        Syed~Muhammad~Talha~Gillani,
        and~Toheed~Omer,%
\thanks{H. Ahmad, S. M. T. Gillani, T. Omer, and S. Aslam are with the
Department of Electrical Engineering, Bahria University, Islamabad,
Pakistan (e-mail: hussainbuic@gmail.com).}%
\thanks{Simulation code (ns-3 module and Python link-level engine) is
available with this article.}}

\markboth{Journal of Lightwave Technology}%
{Ahmad \MakeLowercase{\textit{et al.}}: Diffused-Beam Laser-Diode LiFi Under Realizable Constraints}

\maketitle

\begin{abstract}
Link-budget studies of indoor optical wireless systems frequently assume
receiver parameter sets---large photodetector area, large transimpedance,
and wide bandwidth simultaneously---that violate basic circuit
constraints, and noise budgets that omit the amplifier and laser noise
terms that dominate at exactly those design points. This paper develops a
\emph{realizability-constrained} design-space analysis of a diffused-beam
laser-diode (LD) LiFi link anchored to a hardware prototype built by the
authors. The analysis couples the generalized-Lambertian channel of a
holographic-diffuser source to a receiver model that enforces the
transimpedance-amplifier gain-bandwidth/capacitance constraint and
carries a complete noise budget: shot, feedback-resistor thermal, input
current noise, capacitance-driven voltage-noise gain, and laser relative
intensity noise (RIN). Against this budget we evaluate unipolar $M$-PAM
under two specified FEC tiers (7\%-overhead hard-decision at
$3.8\times10^{-3}$, 20\%-overhead soft-decision at $2\times10^{-2}$),
first-bounce diffuse multipath computed by patch integration, and a
quantitative extended-source eye-safety assessment. The full model
predicts 140~Mb/s net at the prototype's demonstrated 14-m range with
6.7~dB of link margin (OOK, HD tier), 240~Mb/s at the zero-margin
4-PAM/SD reach boundary of 14.0~m, and 480--558~Mb/s at 5~m---a factor 3.9--6.6
below what the same link yields under a naive textbook budget with an
unconstrained receiver, quantifying how strongly idealized assumptions
inflate LiFi rate projections. First-bounce analysis shows the
down-facing-source/up-facing-receiver geometry confines multipath to a
worst-case LOS-to-diffuse ratio of 4.2~dB and delay spreads below
0.13~ns, and the diffused 500-mW source remains a factor $\ge$7.8 under
the Class-1 extended-source emission limit at all viewing distances. All
models are released as an ns-3 module and an independent Python engine
whose mutual agreement is enforced by automated tests; these
predictions are explicitly model-based, and the paper delineates
what the megabit-class prototype does and does not validate.
\end{abstract}


\IEEEpeerreviewmaketitle

\section{Introduction}

\IEEEPARstart{L}{aser} diodes offer indoor optical wireless systems an
order-of-magnitude modulation-bandwidth advantage over phosphor white
LEDs, demonstrated by 9-Gb/s single-channel GaN links~\cite{chi2015gan},
illumination-grade white-laser transmitters~\cite{wu2018white,
chi2017violet}, and real-time Gb/s transceivers~\cite{gao2018real};
LED systems have answered with heavily engineered spectral efficiency,
reaching 15.73~Gb/s with off-the-shelf devices~\cite{bian2019lifi}.
Between these laboratory flagships and deployable LiFi
systems~\cite{haas2016what,haas2020introduction,koonen2018indoor} lies a
less glamorous question: given a \emph{buildable} transmitter and a
\emph{buildable} receiver, what does a diffused-beam LD link actually
support, at what range, under which safety constraints, and with which
honest error bars?

This paper addresses that question quantitatively, and its starting
point is self-critical. An earlier version of this
work~\cite{ahmad2026arxiv} specified the receiver as a large-area
photodiode (1~cm$^2$), a large transimpedance (10~k$\Omega$), and a
wide bandwidth (250~MHz) \emph{simultaneously}, with a noise budget
containing only shot noise and the feedback resistor's $4kT/R_f$.
Such a receiver cannot be built: a 1-cm$^2$ fully-depleted silicon PIN
presents $\sim$35~pF, placing the transimpedance pole near 0.5~MHz,
and no realistic amplifier gain-bandwidth closes the
gap~\cite{sackinger2010limit,mohan2000bw}. The omitted amplifier
terms---input current noise and the capacitance-driven voltage-noise
gain rising as $f^2$, which integrates to
$B^3$~\cite{mohan2000bw}---then dominate the budget at exactly the
wide-bandwidth design points assumed, and laser RIN caps the SNR
exactly where high-order constellations are claimed.

We are deliberately careful about how far this generalizes. The
canonical VLC noise model of~\cite{komine2004fundamental} \emph{does}
carry a capacitance- and transconductance-dependent term, and studies
that adopt it in full inherit that
term~\cite{mach2017d2d}. The pattern we point to is narrower and is
visible in the parameter tables of widely cited work: a 1-cm$^2$
detector tabulated together with 40--200~MHz of bandwidth and a noise
figure given only as a power spectral density, with no amplifier
gain-bandwidth, feedback resistance, or voltage-noise density
stated~\cite{maraqa2024starris,surampudi2018attocell,
tabassum2018coverage}. In network-layer studies such as these the
abstraction is entirely legitimate---the receiver is not the object of
analysis. The failure mode arises when the same convention is carried
into \emph{link-level rate prediction}, as our own earlier
version~\cite{ahmad2026arxiv} did, because the unstated constraint is
then the binding one.

\subsection{Contributions}

\begin{enumerate}
\item \textbf{A realizability-constrained receiver model}
(Section~\ref{sec:receiver}): the gain-bandwidth/capacitance constraint
is enforced explicitly, the étendue trade between collection area and
field of view is carried through the optics, and the noise budget
includes shot, $4kT/R_f$, input current noise $i_n^2B$, the
$(2\pi C_Tv_n)^2B^3/3$ term, and RIN~\cite{sackinger2005broadband,
petermann1988laser}. A design-frontier chart makes infeasible parameter
sets visible at a glance.
\item \textbf{Quantified idealization penalty}
(Section~\ref{sec:results}): at the anchor operating point (14~m,
$20^\circ$ diffuser) the complete budget yields 5.1~dB less SNR than
the naive budget at the same parameters---and 3.9--6.6$\times$ lower
net rate than the earlier unconstrained receiver---providing a
concrete, reproducible measure of how idealized assumptions inflate
projections.
\item \textbf{First-bounce multipath by patch integration}
(Section~\ref{sec:channel}), following~\cite{barry1993simulation,
carruthers1997modeling}: for the ceiling-down/desk-up geometry the
worst-case LOS-to-diffuse power ratio in a $5\times5\times3$-m room is
4.2~dB (wide diffuser, wide FoV, room corner) with first-bounce rms
delay spread $\le0.13$~ns---i.e., $\tau_{\rm rms}R_s<0.02$ at the
150-MBd symbol rate, quantifying rather than assuming the LOS
approximation, within stated limits (first bounce); tilted-receiver
cases are computed and shown to fail on power (outage) before ISI.
\item \textbf{A quantitative extended-source eye-safety assessment}
(Section~\ref{sec:safety}) in the manner of IEC~60825-1~\cite{iec60825}:
the 25-mm, $20^\circ$ holographic diffuser keeps the 7-mm-pupil
collected power a factor $\ge7.8$ below the Class-1 extended-source
limit at every viewing distance from 10~cm, closing the loop between
safety and the link budget that uses the same 450-mW radiated power.
\item \textbf{Specified FEC tiers and modulation}
(Section~\ref{sec:modulation}): unipolar $M$-PAM at symbol rate $=B$
(no subcarrier, hence no double-sideband bandwidth penalty and no
unaccounted bias power split), evaluated against a G.975.1-class
7\%-overhead HD-FEC ($3.8\times10^{-3}$) and a 20\%-overhead SD-FEC
($2\times10^{-2}$)~\cite{itu_g9751}.
\item \textbf{An open, cross-verified simulation framework}
(Section~\ref{sec:simulator}): an ns-3 module (channel, PHY, ARQ MAC,
\texttt{NetDevice}) and an independent Python engine implementing the
same physics, with automated tests enforcing agreement. We are explicit
about epistemic status: this machinery \emph{verifies} the
implementations against each other and against Monte-Carlo statistics;
it does not \emph{validate} the physical assumptions, which only
wideband measurement can do (Section~\ref{sec:scope}).
\end{enumerate}

The study is anchored to the authors' hardware
prototype~\cite{ahmad2019thesis,ahmad2019futuristic}: a 500-mW LD
switched through a CH340 USB--TTL bridge, diffused holographically, and
received by a threshold photodetector over a 14-m indoor LOS path.
(Prior-publication statement: the prototype hardware, its demonstrated
transfers, and the interface-ceiling data of Fig.~13 originate in the
thesis~\cite{ahmad2019thesis}; all modeling, the receiver design, the
simulation framework, and every figure and result in this article are
new.) The
prototype's role is strictly delimited: it establishes end-to-end
feasibility (data, real-time voice, and image transfer) at
interface-limited rates $\le1.59$~Mb/s, and it anchors geometry and
component classes. Every rate above that figure in this paper is a
model prediction, labeled as such, whose experimental confirmation
requires the wideband front end specified in Section~\ref{sec:scope}.

\section{Related Work and Positioning}\label{sec:related}

Every result cited below is labelled \emph{measured} (a physical demonstration), \emph{modelled} (analysis or simulation), or \emph{review}. The convention is deliberate: the gap this paper addresses is the one between what has actually been measured and what indoor link-budget models routinely assume. Broad context for optical wireless in the 6G programme is set by recent surveys and roadmaps \cite{celik2023survey,ata2026survey,chow2024jlt6g,krishnamoorthy2025owc}, which we do not duplicate; we concentrate on the six threads our contributions touch.

\subsection{Laser-Diode Optical Wireless: The Measured State of the Art}
\label{ssec:rel-ld}

At device level the blue GaN laser diode is no longer the limiting element. A c-plane narrow-ridge short-cavity mini-LD has been measured with a $-3$\,dB modulation bandwidth of 5.9\,GHz and a 1.02\,W/A slope efficiency, supporting 20.06\,Gb/s with bit- and power-loaded DMT \cite{wang2024gan5ghz} (measured); InGaN quantum barriers push the intrinsic bandwidth beyond 8\,GHz at 1.85\,W/A slope efficiency and 29\% wall-plug efficiency, with 36.5\,Gb/s reported \cite{jia2025blueld} (measured); and an InGaN mini-LD with 8.4\,GHz measured bandwidth reaches 33\,Gb/s using PAM-4 rather than multicarrier modulation \cite{hu2026minild} (measured). The underlying epitaxial and cavity-geometry trends are consolidated in two recent reviews \cite{gu2026gan,hou2024tutorial}.

System-level records, however, are set in geometries very unlike a room. A ten-wavelength packaged module reached 113.175\,Gb/s with a sparse-structured reservoir-computing post-equaliser \cite{luo2024rainbow}; a 50-channel WDM intra-connect reached 534.51\,Gb/s \cite{chi2025wdm500}; and 601.46\,Gb/s has been reported over a 1\,m MMF--1\,m free-space--1\,m MMF path with constellations up to 1024-QAM \cite{zhou2025col600} (all measured). Underwater, 170.1\,Gb/s was obtained with a five-wavelength module and a reciprocal differential receiver whose learned equaliser is itself worth 16.1\,Gb/s \cite{lu2025underwater} (measured). The closest prior art to an illuminating laser luminaire is the laser-based SMD white source of \cite{chen2024jlt100g}, which delivers 450\,lm at over 1000\,cd/mm$^2$ brightness while a ten-channel WDM link exceeds 100\,Gb/s aggregate with Volterra equalisation, plus 4.8\,Gb/s over a 500\,m outdoor point-to-point path (measured). These are short, aligned or guided paths sustained by heavy digital signal processing; none is a room-scale broadcast link operated under an eye-safety cap. That regime, not the record table, is the subject of this paper.

\subsection{Eye Safety, Speckle, and the Diffused-Source Premise}
\label{ssec:rel-safety}

The diffused-beam transmitter has an explicit experimental precedent: a phosphorous diffuser bonded to a blue GaN laser diode both broadens the luminescent spectrum and diverges the beam spot for room illumination while carrying 5.2\,Gb/s 16-QAM OFDM over 60\,cm at BER $3.1\times10^{-3}$ \cite{chi2015diffuser} (measured). Laser-pumped phosphor lighting is itself a reviewed, commercially deployed source class \cite{rahman2022laserphosphor}. The two standing objections to laser illumination have been quantified and answered: additively blended RGB laser white achieves speckle contrast $\le 5\%$ at 150\,lm/W luminous efficacy while carrying multi-Gb/s data \cite{lazzaro2026speckle} (measured), cascaded diffusing stages have been measured to suppress speckle contrast from 0.32 (stationary diffuser) through 0.14 and 0.09 to 0.05 for a complete module with phosphor plate \cite{kumar2024speckle} (measured), and the design variables that control residual speckle in a phosphor-converted laser source have been isolated systematically \cite{jensen2025speckle} (measured).

For safety, the tutorial treatment of IEC 60825-1 for optical wireless links \cite{soltani2022safety} (review) establishes the extended- versus point-source distinction, apparent source size, and the exposure limits that bound transmit power. The specific classification machinery our analysis uses --- characterising the apparent-source angular subtense, scaling the emission limit by $C_6$ once the source exceeds the minimum angle, and classifying at the ``most restrictive position'' rather than the closest one --- is set out in \cite{schulmeister2015extended} (modelled), which also warns that a product need not have a single unique apparent source. A committee-side account of the standard's development \cite{wheatley2026iec} (review) documents the difficulty of keeping classification aligned with fast-moving source technology, which is why we present our margin as conditional on a diffuser-integrity interlock rather than as a settled classification.

\subsection{LED Systems and the Matched-Condition Comparison}
\label{ssec:rel-led}

Any laser-versus-LED claim must be stated at matched distance, matched emitter count, and matched receiver-side processing, or it is not a comparison at all. The canonical deployable-LED benchmark remains 15.73\,Gb/s over 1.6\,m from four off-the-shelf LEDs with OFDM and adaptive bit loading \cite{bian2019lifi} (measured). The headline aggregate figure, 25.20\,Gb/s, uses eight emitters spanning 276--655\,nm over 25\,cm; the visible-band-only portion is 20.11\,Gb/s \cite{qiu2022beyond25} (measured). An all-visible RGBP module reaches 23.43\,Gb/s over 1\,m, but only after transmitter-side pre-equalisation raises per-channel bandwidths to 756.6--929\,MHz \cite{tang2023rgbp} (measured). At 2\,m, 16.6\,Gb/s is obtained from tricolour mini-LEDs with per-colour bandwidths of 668--859\,MHz and a five-layer neural-network receiver replacing conventional estimation, equalisation and demodulation \cite{liu2021sdmwdm} (measured).

Room-scale LED evidence is thinner and more instructive. A $3\times3$ violet micro-LED array with 13.4\,mW transmitted power and only 350\,MHz of device bandwidth delivers 10.23, 10.10 and 9.51\,Gb/s at 0.2, 1 and 10\,m below the $3.8\times10^{-3}$ threshold, using bit loading with distance-adaptive pre-equalisation \cite{jin2023tenm} (measured); a single 60\,$\mu$m c-plane micro-LED reaches 10.547\,Gb/s at the optimum of an emitter-size sweep and still delivers 8.649\,Gb/s at 10\,m \cite{rao2024cplane} (measured). Elsewhere the rate is bought with processing or with bespoke devices: 8.75\,Gb/s from a 75\,$\mu$m blue micro-LED of 1025\,MHz bandwidth using an artificial neural-network equaliser \cite{wei2021ann}, 19.3\,Gb/s over 1.2\,m from a spatially multiplexed array at over 9.5\,Gb/s per emitter \cite{jin2025sdm}, and 15.64\,Gb/s at 0.3\,m using a red GaN micro-LED \emph{as the photodetector} \cite{ai2025microled} (all measured). Adaptive loading with an explicit SNR gap \cite{jin2025green} (measured) and physics-informed reservoir-computing equalisation, worth roughly a 10\% rate gain with an SNR advantage of 1.2\,dB over a Volterra equaliser and 1.5\,dB over conventional reservoir computing \cite{dong2026rc} (both), show that LED baselines are not static. At the opposite extreme of emitter area, RGB-single-chip OLEDs top out at 3.2\,Gb/s with $-6$\,dB bandwidths of 171--255\,MHz \cite{yoshida2024oled} (measured). We therefore report no rate record and make no laser-beats-LED claim; we compare architectures at stated distance and stated receiver complexity.

\subsection{Receiver-Side Limits: Where Our Frontier Sits}
\label{ssec:rel-rx}

That receiver aperture cannot be increased for free is established prior art, not our contribution. The rate--FOV trade-off arising from the compounding of photodetector area--bandwidth and optical gain--FOV constraints is derived in \cite{soltani2023imaging} (modelled), where an ``array of arrays'' imaging receiver of $2\times2$\,cm is optimised to roughly 24\,Gbps at 15$^\circ$ FOV; the non-imaging counterpart optimises multi-tier angle-diversity receivers under both constraints jointly and reaches 12\,Gb/s at 30$^\circ$ half-angle over a 3\,m laser link with height $\le0.5$\,cm and effective area $\le0.5$\,cm$^2$ \cite{sarbazi2024receivers} (modelled). The information-theoretic statement that an interior optimum detector area exists --- too large costs bandwidth, too small costs captured power --- is given in \cite{bashir2024pdsize} (modelled), and the practice of treating the front end as a frequency-dependent gain-to-noise profile modelled by a pole--zero transfer function, rather than a single $3$\,dB number, is developed in \cite{liu2026gnr} (both).

The known escape routes each substitute one ceiling for another. Luminescent solar concentrators evade the etendue limit but are shown analytically to be equivalent to an RC low-pass network, with fluorescence lifetime and self-reabsorption as the dominant bottlenecks \cite{portnoi2021lsc} (both); $50\times50$\,mm luminescent antennas nonetheless outperform a bare photodiode under white light \cite{meucci2024lsc} (both), three such antennas benchmarked under both laser and LED sources reach above 70\,Mb/s with NRZ-OOK \cite{meucci2025comparative} (measured), and a fluorescent slab of 245\,MHz bandwidth coupled to a SiPM yields 1.4\,Gb/s over 30\,cm under 500\,lux with 20-fold ambient rejection \cite{ali2022sipm} (measured). Segmentation is an active route: an APD array laid out to match a fluorescent antenna's side-emission profile achieves higher power gain at comparable bandwidth to a single detector of nearly the same total area \cite{wang2026apdarray} (both). Photon-counting front ends trade bandwidth for sensitivity --- 30\,pW, or 41.5 photons per bit, at 1\,Mbit/s \cite{liu2023uvsipm} (measured) --- with nonlinearity and pulse width as the binding limits \cite{matthews2023roadmap} (modelled) and dead-time blocking and ISI as the PAM-specific failure mode \cite{wang2026spad} (modelled). Several of these works do carry front-end effects into the achievable rate, notably the jointly constrained optimisation of \cite{sarbazi2024receivers} and the pole--zero gain-to-noise treatment of \cite{liu2026gnr}.

The amplifier side of the problem is, separately, a mature body of theory that the optical-wireless modelling literature does not always meet. The bound tying transimpedance to bandwidth through the amplifier gain--bandwidth product and the total input capacitance is the \emph{transimpedance limit} \cite{sackinger2010limit}, treated at book length in \cite{sackinger2017tia} and derived in equivalent form as $R_T\le A\omega_A/(C_T\omega_{3\mathrm{dB}}^2)$ in \cite{mohan2000bw} (measured), whose CMOS front end also states explicitly that the dominant input-referred noise densities rise as $f^2$ and therefore dominate once integrated --- the $B^3$ behaviour our budget carries. The lineage runs back to the classical optical-receiver analyses of \cite{personick1973rx,hullett1976tia,muoi1984rx}, and VLC-specific front-end work confronts the same trade directly, whether by regulated-cascode topologies sized for high-capacitance photodiodes \cite{kassem2019tia} (measured) or by bandwidth-extension techniques for visible-light receivers \cite{cura2013tia} (measured). We therefore claim no circuit-theoretic novelty whatever. What we add is narrower and lies at the seam between the two literatures: tying the \emph{assumed} detector area of a link-budget study to a specific gain--bandwidth product and feedback resistance through the implied junction capacitance, and propagating the resulting voltage-noise term into reach and rate, so that a parameter set which cannot be built is visible as such before any rate is quoted.

\subsection{Modulation and Coding for IM/DD}
\label{ssec:rel-mod}

The choice of unipolar $M$-PAM over optical OFDM is supported rather than asserted. For indoor multipath IM/DD links under an average-optical-power constraint, unipolar $M$-PAM with MMSE decision-feedback equalisation is the most power-efficient scheme over the practical range, with OFDM variants requiring several dB more average optical power at equal spectral efficiency \cite{barros2012ofdmpam} (modelled). The normalised-power-budget framing of \cite{mardanikorani2020pam} (modelled) makes the boundary quantitative: optical OFDM becomes advantageous only above roughly a 60\,dB power budget, DCO-OFDM beats PAM in the 30--60\,dB range only with adaptive bit loading, without bit loading PAM is about 2.5\,dB better at 60\,dB, and below 30\,dB PAM is always preferred. Capacity bounds for the non-negativity-constrained Gaussian IM/DD channel under average- and peak-power constraints \cite{chaaban2022capacity} (review) supply the correct ceiling against which unipolar rates should be reported. Emitter-side clipping and static nonlinearity, which penalise high-PAPR waveforms in a peak-constrained source, are quantified in \cite{deng2018lednl} (both), and multilevel PAM with equalisation is what practice actually uses to reach multi-Gb/s: 8.8\,Gb/s PAM-4 with an external modulator and a neural-network equaliser \cite{shi2023pam4} (measured).

Laser RIN is the term most often omitted. For LD-based indoor VLC, RIN emerges as the dominant noise source once transmitted optical power is raised \cite{yaseen2025rin} (modelled), a conclusion extended to random channels in \cite{elfar2026vlcnoise} (modelled). Because RIN is signal-dependent, the signal-independent Gaussian error formula is not exact, and a closed-form SER for the optimal RIN-aware detector making no assumption on constellation geometry is available \cite{villenas2025rin} (modelled). The magnitude matters for visible GaN devices specifically: mode-hopping and mode-competition dynamics drive InGaN edge-emitter RIN roughly 9\,dB above comparable AlGaAs lasers \cite{congar2018rin} (measured), which is why we sweep RIN rather than adopt a telecom value, and why our 16-PAM tier is flagged as RIN-fragile. Shaping is the natural next step but is out of scope here; we note that probabilistically shaped unipolar 4-PAM under an explicit peak-power constraint is compatible with a PAS/FEC chain \cite{wiegart2021ps4pam} (both), that per-subcarrier shaping has measured 10.81\,Gb/s in a WDM VLC link, 25\% above adaptive bit-and-power loading under identical channel conditions \cite{gutema2022ps} (measured), and that probabilistic shaping has been benchmarked against geometric shaping for bandwidth-limited VLC \cite{kafizov2024pcs} (modelled).

We report rates at two specified FEC tiers rather than at an unlabelled ``FEC limit''. The 7\% hard-decision and $\sim$20\% soft-decision overhead classes, with their net coding gains and input BER thresholds, come from the 100G transport literature \cite{chang2010fec} (review). We also state the caveat: a channel-independent soft-decision FEC threshold is invalid for soft-decision bit-wise decoding and can underestimate spectral efficiency by up to 20\% at low code rates, with generalised mutual information the consistent predictor \cite{alvarado2016fec} (both). The conditions under which quoting uncoded BER against a threshold is legitimate, and the mismatched-receiver alternative, are given as implementable recipes in \cite{agrell2021recipes} (review); our claims ledger follows that discipline.

\subsection{Indoor Channel Modelling: Multipath, Mobility, Orientation}
\label{ssec:rel-chan}

The taxonomy of indoor VLC channel-modelling methods, from recursive and Monte Carlo approaches to non-sequential ray tracing, is surveyed in \cite{miramirkhani2020channel} (review), and the ray-traced reference impulse responses adopted by IEEE 802.11bb for industrial, medical, enterprise and residential usage models are documented in \cite{miramirkhani2023bb} (modelled). Published delay-spread magnitudes bound what any first-bounce treatment must be consistent with: an industrial ray-tracing study reports RMS delay spreads of 6--11\,ns under a single transmitter, rising to means of 13.79\,ns and 14.78\,ns at 1.5\,m and 2.5\,m receiver height under multiple transmitters \cite{tong2023industrial} (modelled), while a 3D non-stationary geometry-based stochastic model validates its simulated 3\,dB channel bandwidth against existing measurement data \cite{zhu2022gbsm} (both). Our truncation after the first bounce is a stated approximation, and the literature indicates its direction of error rather than excusing it: the measurement-validated hybrid of \cite{mana2021multilink} (both) treats early reflections explicitly in the frequency domain \emph{and} adds an integrating-sphere term for higher orders, precisely because truncation alone is not sufficient in the general case; its sub-5\% relative mean square error is reported for the LOS component. Higher-order bounces therefore add a weaker, longer-delayed tail that our first-bounce figures do not capture, and we treat our delay-spread numbers as a lower bound whose margin against the symbol period (below 2\% at 150\,MBd) is large enough to absorb a several-fold increase. The regime in which the diffuse component stops being a correction at all --- blocked LOS and large access-point separation --- is characterised in \cite{chen2021uplink} (modelled).

Our fixed-normal geometry is a stated scope limit, not an oversight. Measured device orientation from 40 participants is well fitted by a Laplace distribution for static users and a Gaussian for mobile users \cite{soltani2019orientation} (measured), with closed-form channel-gain statistics (modified truncated Laplace, modified Beta, and their sums) derived in \cite{arfaoui2021channel} (both); these supply the distributions for a stochastic-orientation extension. Measured rotation and movement traces put the LiFi channel coherence time on the order of tens of milliseconds and show that mobile operation requires at least two differently oriented photodiodes \cite{ma2024mobile} (both), which bounds the interval over which our static link budget is valid.
\begin{table*}[!t]
\centering
\caption{Positioning against the strongest recent \emph{measured} demonstrations. The final column is the wedge: what each source reports about the analog receiver front end.}
\label{tab:related}
\scriptsize
\renewcommand{\arraystretch}{1.15}
\begin{tabular}{@{}p{0.13\textwidth} p{0.19\textwidth} p{0.20\textwidth} p{0.11\textwidth} p{0.28\textwidth}@{}}
\toprule
\textbf{Work} & \textbf{Emitter / scheme} & \textbf{Rate / reach} & \textbf{Evidence} & \textbf{Receiver detail reported} \\
\midrule
Zhou \emph{et al.} \cite{zhou2025col600} & 50-ch WDM LD, bit/power-loaded DMT, up to 1024-QAM & 601.46\,Gb/s over 1\,m MMF + 1\,m free space + 1\,m MMF & Measured & Differential pilot coding; no front-end bandwidth or noise budget stated \\
Chen \emph{et al.} \cite{chen2024jlt100g} & Laser-based white SMD source, 10-ch WDM & $>$100\,Gb/s aggregate indoor; 4.8\,Gb/s over 500\,m outdoor & Measured & Volterra nonlinear equalisation; illumination reported (450\,lm, $>$1000\,cd/mm$^2$); front end not stated \\
Hu \emph{et al.} \cite{hu2026minild} & InGaN mini-LD, PAM-4 & 33\,Gb/s; 8.4\,GHz measured device bandwidth & Measured & Emitter bandwidth reported; receiver front end not stated \\
Chi \emph{et al.} \cite{chi2015diffuser} & Blue LD + phosphorous diffuser (diverged beam) & 5.2\,Gb/s 16-QAM OFDM over 0.6\,m, BER $3.1\times10^{-3}$ & Measured & Not stated \\
Lazzaro \emph{et al.} \cite{lazzaro2026speckle} & Blended R/G/B laser white light & Multi-Gb/s; speckle contrast $\le5\%$; 150\,lm/W & Measured & Illumination metrics reported; front end not stated \\
Bian \emph{et al.} \cite{bian2019lifi} & 4 off-the-shelf LEDs, WDM OFDM, adaptive bit loading & 15.73\,Gb/s over 1.6\,m & Measured & Not stated \\
Jin \emph{et al.} \cite{jin2023tenm} & $3\times3$ violet micro-LED array, 13.4\,mW, 350\,MHz & 10.23 / 10.10 / 9.51\,Gb/s at 0.2 / 1 / 10\,m & Measured & Distance-adaptive pre-equalisation (Tx side); receiver noise not stated \\
Ai \emph{et al.} \cite{ai2025microled} & Red GaN micro-LED used \emph{as} the photodetector & 15.64\,Gb/s over 0.3\,m & Measured & Detector device characterised; bespoke, non-commodity front end \\
Ali \emph{et al.} \cite{ali2022sipm} & 405\,nm laser Tx; fluorescent antenna + SiPM & 1.4\,Gb/s over 0.3\,m under 500\,lux & Measured & Fluorophore bandwidth 245\,MHz; 20$\times$ ambient rejection (200$\times$ filtered) \\
\textbf{This work} & Diffused-beam LD luminaire, unipolar $M$-PAM / OOK & 558\,Mb/s @ 5\,m; 279\,Mb/s @ 12\,m; 140\,Mb/s @ 14\,m; OOK reach 20.7\,m (HD) / 23.6\,m (SD) & Computed, cross-verified; anchored to a 14\,m UART-limited (1.59\,Mb/s) prototype & TIA closed-loop $B\!\approx\!\sqrt{\mathrm{GBW}/(2\pi R_f C_T)}$ frontier vs.\ PD area/capacitance; five-term noise budget (shot, $4kT/R_f$, $i_n^2B$, $(2\pi C_T v_n)^2B^3/3$, RIN) \\
\bottomrule
\end{tabular}
\end{table*}
\subsection{Networking, Standards, and Simulation Tooling}
\label{ssec:rel-net}

Two IEEE light-communication standards are directly relevant. IEEE 802.11bb reuses the 802.11 MAC and OFDM-based PHYs over optical links in the 800--1000\,nm band with MAC-SAP throughput specified from 10\,Mb/s to 9.6\,Gb/s, and its stated open problems include interoperability across solid-state sources with differing modulation bandwidths \cite{khorov2022bb} (review), with its reference channels in \cite{miramirkhani2023bb}. That open problem concerns the \emph{emitter} side; the constraint we add is its receiver-side counterpart, and the two together bound what a heterogeneous light-communication deployment can assume of an arbitrary endpoint. IEEE Std 802.15.13 specifies a deterministic dynamic-TDMA MAC with two PHYs and has been implemented and experimentally tested for industrial use \cite{bober2024ojvt} (measured), and ITU-T G.9991 specifies a third track for in-premises optical networking \cite{itu_g9991}. Our rates fall inside the 802.11bb MAC-SAP envelope, and we report them against that envelope rather than against interconnect records.

The attocell abstraction and its challenge list --- inter-cell interference, uplink asymmetry, mobility and handover --- are defined in \cite{haas2020indoor} (review), with the hybrid LiFi/WiFi context in \cite{wu2021hybrid} (review), the laser-based ``LiFi 2.0'' roadmap that names receiver front-end design as a limiting subsystem in \cite{soltani2023lifi2} (review), and the IoT deployment pull in \cite{linnartz2022eliot} (review). Learning-based load balancing and mobility management define the current network-layer baseline \cite{ji2024atcnn,ji2025usercentric} (modelled), handover thresholds keyed to optical gain and incidence angle are given in \cite{murad2022handover} (modelled), and a COTS LiFi/WiFi testbed with conveyor-belt-emulated motion reduces handover outage to under one second at 20\,Mb/s \cite{ancillotti2025cots} (measured) --- a useful reminder of the distance between standardised capability and deployed behaviour.

On tooling, the precedent for open-source ns-3 optical modules is \cite{aldalbahi2017ns3} (both), the first such VLC module, validated against a software-defined-radio testbed with phosphor-converted white LEDs. The closest existing artefact is the ns-3 LiFi framework of \cite{ullah2021ns3} (modelled), which already implements LOS plus first-order NLOS reflections, SNR/BER for OOK, PAM and QAM, orientation-aware mobility, a TDMA MAC and vertical handover, with source released. We state the delta plainly: that framework models neither a laser-diode source with RIN, nor a TIA-limited receiver whose bandwidth is tied to photodiode capacitance, nor the amplifier voltage- and current-noise terms that dominate our budget. Our module is an extension of that modelling layer, not a replacement for it.

\subsection{Positioning}
\label{ssec:rel-pos}

Table~\ref{tab:related} places this work against the strongest recent measured demonstrations. We claim no rate record and no new device. What we add is narrower and, we hope, more durable: (i) a receiver realizability constraint that ties assumed bandwidth to a specific TIA gain--bandwidth product, feedback resistance and photodiode capacitance. That the constraint bites is not hypothetical: an earlier version of this work~\cite{ahmad2026arxiv} assumed a 1\,cm$^2$ detector, a 10\,k$\Omega$ feedback resistance and 250\,MHz of bandwidth simultaneously, a combination the frontier of Section~\ref{sec:receiver} shows to be unbuildable by a factor above five, and on which its headline rates depended. We correct that here rather than defend it, and we note that the receiver column of Table~\ref{tab:related} shows how commonly the analog front end is simply not reported, which is what allows such combinations to pass unexamined. (ii) A noise budget in which the amplifier terms, including the capacitance-driven $(2\pi C_T v_n)^2B^3/3$ contribution, account for roughly 69\% of the variance at our anchor point, so that a shot-plus-thermal budget flatters SNR by 5.1\,dB. (iii) Rates reported at two named FEC tiers under unipolar $M$-PAM, following the reporting discipline of \cite{agrell2021recipes,alvarado2016fec}. (iv) A first-bounce diffuse analysis giving a worst-case LOS/diffuse ratio of 4.2\,dB and $\tau_{\mathrm{rms}}\le0.13$\,ns, from which tilted receivers are shown to fail on power before ISI; higher-order bounces are not modelled, and \cite{mana2021multilink} indicates the form the residual takes. (v) An eye-safety margin computed with the extended-source machinery of \cite{schulmeister2015extended,soltani2022safety}, stated as conditional on a diffuser-integrity interlock. (vi) Two independent implementations cross-verified against each other, with every number labelled measured, computed or predicted. Cross-verification is not validation, and we do not describe it as such.

\section{System Architecture}\label{sec:architecture}

\begin{figure*}[!t]
\centering
\resizebox{\textwidth}{!}{%
\begin{tikzpicture}[
  node distance=3.2mm and 4.5mm,
  blk/.style={draw, rounded corners=1pt, minimum height=6.5mm,
              minimum width=15mm, align=center, font=\scriptsize},
  opt/.style={blk, fill=blue!8},
  ele/.style={blk, fill=orange!10},
  net/.style={blk, fill=green!8},
  lbl/.style={font=\scriptsize\itshape},
  arr/.style={-{Stealth[length=1.6mm]}, semithick}]
\node[net] (src) {Data\\ source};
\node[ele, right=of src] (mod) {Unipolar M-PAM\\ + FEC encoder};
\node[ele, right=of mod] (drv) {Linear LD\\ driver};
\node[opt, right=of drv] (ld) {500-mW LD\\ (650 nm)};
\node[opt, right=of ld] (dif) {Holographic\\ diffuser $\Phalf$, $T_d$};
\node[blk, right=of dif, fill=gray!12, minimum width=17mm]
  (ch) {LOS + first-bounce\\ channel};
\node[opt, right=of ch] (pd) {7-mm$^2$ PIN +\\ lens/concentrator};
\node[ele, right=of pd] (tia) {TIA $R_f$, $C_T$\\ $i_n$, $v_n$};
\node[ele, right=of tia] (dem) {PAM decision\\ + FEC decoder};
\node[net, right=of dem] (snk) {Data\\ sink};
\draw[arr] (src) -- (mod);
\draw[arr] (mod) -- (drv);
\draw[arr] (drv) -- (ld);
\draw[arr] (ld) -- (dif);
\draw[arr, dashed] (dif) -- (ch);
\draw[arr, dashed] (ch) -- (pd);
\draw[arr] (pd) -- (tia);
\draw[arr] (tia) -- (dem);
\draw[arr] (dem) -- (snk);
\node[lbl, below=1.2mm of ch] {optical, free space};
\end{tikzpicture}%
}
\caption{Modeled transceiver chain. The hardware prototype realized this
chain with a CH340 UART driver and threshold receiver (megabit class);
the model assumes the wideband linear chain whose realizability
Section~\ref{sec:receiver} enforces. The return direction for
acknowledgements is a symmetric link on a separated wavelength
(Section~\ref{sec:simulator}).}
\label{fig:arch}
\end{figure*}
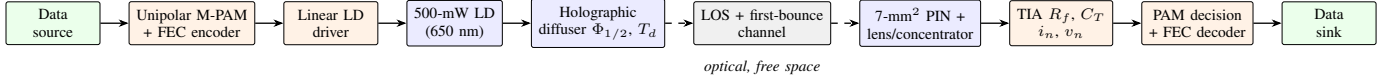

Fig.~\ref{fig:arch} shows the modeled chain. Two receiver profiles span
the terminal design space:

\emph{Desk profile} (default): a fixed terminal aimed at the ceiling
luminaire; field-of-view semi-angle $\Psi_c=15^\circ$, which lets the
étendue-limited optics concentrate $g=n^2/\sin^2\Psi_c=33.6$ onto the
small photodiode, and a 10-nm interference filter with narrow FoV
suppressing ambient photocurrent to $I_{bg}=20~\mu$A. \emph{Pointing
budget}: inside the FoV cone the $\cos\psi$ penalty is benign
($\le$0.2~dB at $\pm12^\circ$, 0.3~dB at the $15^\circ$ edge), so the
requirement is capture within a $\pm15^\circ$ cone---coarse aiming for
a fixed terminal---while the FoV boundary itself is a hard cutoff;
the aligned-boresight numbers quoted later carry this stated
tolerance.

\emph{Mobile profile}: $\Psi_c=60^\circ$ ($g=3.0$), $I_{bg}=100~\mu$A;
used for room-coverage analysis under a lamp.

The prototype~\cite{ahmad2019thesis} anchors the transmit side
(500-mW-class LD, 25-mm holographic diffuser, 14-m indoor path) and the
demonstrated payloads; its CH340 interface bounds its measurable
goodput at 1.59~Mb/s (Section~\ref{sec:scope}), and its threshold
receiver anchors no parameter of the wideband receiver model, which is
specified from component classes in Section~\ref{sec:receiver}.

\section{Optical Channel: LOS and First-Bounce Multipath}
\label{sec:channel}

\subsection{LOS Gain}

The diffused emission is modeled as a generalized Lambertian lobe of
order $m=-\ln 2/\ln\cos\Phalf$; the diffuser transmission $T_d=0.9$ is
carried explicitly, so the radiated power is $P_t=T_dP_{\rm LD}=450$~mW.
For detector area $A$, filter transmission $T_s$, concentrator gain
$g(\psi)$, emission angle $\phi$ and incidence $\psi\le\Psi_c$
\cite{kahn1997wireless,komine2004fundamental}:
\begin{equation}
H_{\rm LOS}=\frac{(m+1)A}{2\pi d^{2}}\cos^{m}\!\phi\;T_s\,g(\psi)\cos\psi .
\label{eq:H0}
\end{equation}

\subsection{First-Bounce Diffuse Component}

\begin{figure*}[!t]
\centering
\includegraphics[width=\textwidth]{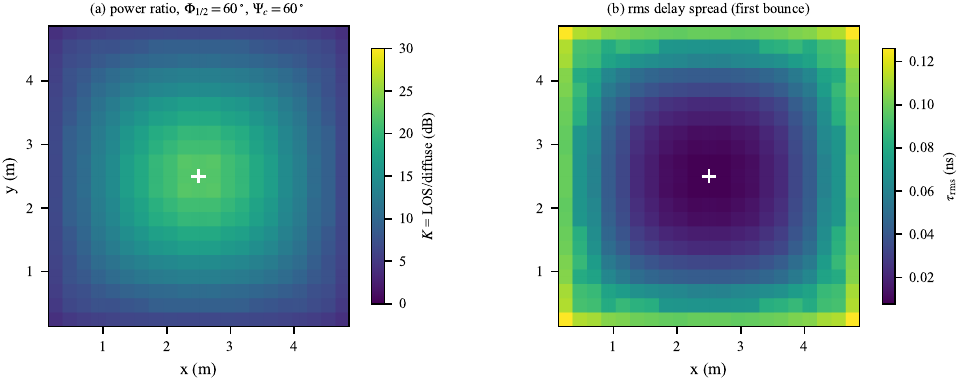}
\caption{First-bounce multipath over the desk plane of a
$5\times5\times3$-m room (source: ceiling center, $\Phalf=60^\circ$;
receiver: up-facing, $\Psi_c=60^\circ$; $\rho=0.7$): (a) LOS-to-diffuse
power ratio $K$; (b) rms delay spread. Worst case $K=4.2$~dB at the
room corner with $\tau_{\rm rms}=0.13$~ns, i.e.\
$\tau_{\rm rms}R_s<0.02$ at 150~MBd.}
\label{fig:multipath}
\end{figure*}

Reviewer-facing honesty requires the LOS assumption to be quantified,
not asserted. We compute the first-bounce response by patch integration
over the walls and ceiling ($\rho=0.7$, 25-cm patches)
following~\cite{barry1993simulation}: each patch receives
$(m{+}1)\cos^m\phi_1\cos\theta_1/2\pi d_1^2$ from the source and
re-radiates as a first-order Lambertian toward the receiver, subject to
the receiver FoV. Fig.~\ref{fig:multipath} maps the resulting
LOS-to-diffuse power ratio $K$ and rms delay spread $\tau_{\rm rms}$
(power-weighted, LOS included) across the desk plane for the widest
configuration (source $\Phalf=60^\circ$, mobile receiver
$\Psi_c=60^\circ$)---the case most favorable to multipath.

Three conclusions. (i) Even in the widest case, first-bounce power
remains 4.2~dB below LOS at the worst corner and $>$15~dB below over
the central two-thirds of the room; for the desk profile
($\Psi_c=15^\circ$) no first-bounce path falls inside the FoV at all
(floor bounces arrive from below an up-facing detector). The physical
reason is geometric: a down-pointing source illuminates predominantly
the floor, whose bounce cannot couple into an up-facing receiver.
(ii) Delay spread stays below 0.13~ns, so intersymbol interference at
the 150-MBd symbol rate is negligible ($\tau_{\rm rms}R_s<0.02$;
$<$0.2 even if higher-order bounces multiplied $\tau_{\rm rms}$
several-fold). (iii) The diffuse term slightly \emph{aids} corner
coverage power. (iv) \emph{Tilted receivers}, evaluated with the same
patch model: a corner terminal tilted $45^\circ$ \emph{toward} the
luminaire improves $K$ to 7.1~dB; tilted $30$--$45^\circ$ \emph{away},
the luminaire leaves the FoV entirely and the link becomes
diffuse-only---received power then falls below the OOK threshold long
before ISI matters (diffuse-only $\tau_{\rm rms}\le0.34$~ns,
$\tau_{\rm rms}R_s\le0.05$). Tilt is therefore a coverage/outage
phenomenon to be handled by handover at network level, not an ISI
regime. Scope: first bounce only; higher-order bounces add a weaker,
longer tail and are the remaining extension of the released
channel-model class.

\section{Receiver Realizability and the Complete Noise Budget}
\label{sec:receiver}

\subsection{The Gain-Bandwidth/Capacitance Constraint}

\begin{figure*}[!t]
\centering
\includegraphics[width=\textwidth]{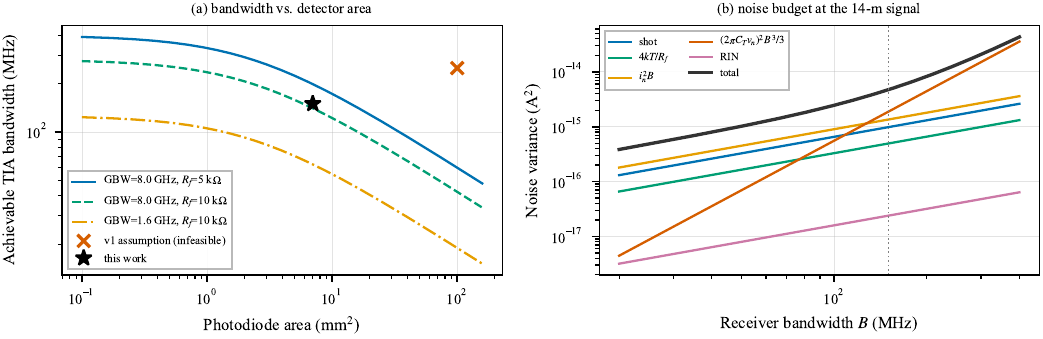}
\caption{(a) TIA-achievable bandwidth versus photodiode area
(\eqref{eq:tia}, 150-$\mu$m PIN): the v1 assumption sits far above
every frontier; the revised design point lies on the 8-GHz/5-k$\Omega$
frontier with margin. (b) Complete noise budget versus bandwidth at the
14-m received signal: above $\sim$120~MHz the capacitance-driven
$(2\pi C_Tv_n)^2B^3/3$ term dominates; the naive shot$+4kT/R_f$ budget
(bottom two curves) captures less than a third of the total at
150~MHz.}
\label{fig:receiver}
\end{figure*}

A transimpedance stage of feedback resistance $R_f$ over total input
capacitance $C_T$ (photodiode $+$ amplifier input) built on an
amplifier of gain-bandwidth product GBW achieves closed-loop
bandwidth~\cite{sackinger2010limit,sackinger2017tia,mohan2000bw}
\begin{equation}
B \approx \sqrt{\frac{\mathrm{GBW}}{2\pi R_f C_T}} .
\label{eq:tia}
\end{equation}
We claim no novelty for \eqref{eq:tia}: it is the transimpedance limit
of the broadband-circuits literature~\cite{sackinger2010limit}, the
same bound that appears as $R_T\le A\omega_A/(C_T\omega_{3\rm
dB}^2)$ in~\cite{mohan2000bw} and descends from the classical
optical-receiver analyses of~\cite{personick1973rx,hullett1976tia,
muoi1984rx}. What is novel here is only its use as a \emph{binding
admissibility test on a published link budget}, applied before any
rate is quoted. VLC-specific front ends that confront the same
constraint---regulated-cascode topologies for high-capacitance
photodiodes~\cite{kassem2019tia} and bandwidth-extension techniques
for visible-light receivers~\cite{cura2013tia}---show that the
trade is live in this field and not merely textbook.
Fig.~\ref{fig:receiver}(a) plots this frontier against photodiode area
(fully depleted 150-$\mu$m PIN, $C_d=\varepsilon_{\rm Si}A/w$). The
parameter set assumed in the earlier version of this
work---$A=1$~cm$^2$, $R_f=10$~k$\Omega$, $B=250$~MHz---sits a factor
$>$5 above the frontier even at $\mathrm{GBW}=8$~GHz and is
unrealizable; this error class is common in link-budget studies and is
exactly what the frontier chart exposes. The revised design point,
$A=7$~mm$^2$ ($C_d=4.9$~pF, $C_T=6.5$~pF), $R_f=5$~k$\Omega$,
$B=150$~MHz, requires $\mathrm{GBW}=2\pi R_fC_TB^2=4.6$~GHz---within
reach of modern wideband amplifiers---and recovers collection area
optically: the étendue-limited concentrator of the $15^\circ$ desk
profile yields an effective aperture $A\,g=235$~mm$^2$.

\subsection{Noise Budget}

With photocurrent $I_{ph}=\mathcal{R}P_r$ and background current
$I_{bg}$, the input-referred noise over bandwidth $B$
is~\cite{kahn1997wireless,petermann1988laser,mohan2000bw}\cite[Chs.~4--5]{sackinger2017tia}
\begin{align}
\sigma^{2} ={}& 2q(I_{ph}{+}I_{bg})B \;+\; \frac{4k_BT}{R_f}B
\;+\; i_n^{2}B \nonumber\\
&{}+ \frac{(2\pi C_T v_n)^{2}}{3}B^{3}
\;+\; \mathrm{RIN}\cdot I_{ph}^{2}B ,
\label{eq:noise}
\end{align}
with amplifier input noise densities $i_n=3$~pA/$\sqrt{\rm Hz}$,
$v_n=1$~nV/$\sqrt{\rm Hz}$ (representative of wideband TIAs), and laser
$\mathrm{RIN}=-120$~dB/Hz as the default for a multimode visible LD,
swept over $-110$ to $-130$~dB/Hz in Section~\ref{sec:results} because
vendor data for low-cost visible LDs is scarce---an uncertainty we
propagate rather than hide. The RIN term imposes a distance-independent
SNR ceiling $1/(\mathrm{RIN}\cdot B)$: 38.2~dB at $-120$~dB/Hz but only
28.2~dB at $-110$~dB/Hz, which brackets the 16-PAM thresholds and is
why the highest-order results carry an explicit RIN sensitivity figure.
Fig.~\ref{fig:receiver}(b) shows the budget at the 14-m signal level:
at $B=150$~MHz the amplifier terms contribute 69\% of the total
variance and the naive budget overstates SNR by 5.1~dB.
Table~\ref{tab:params} lists all parameters.

\begin{table}[!t]
\caption{System Parameters (Revised, Realizability-Checked)}
\label{tab:params}
\centering
\begin{tabular}{lll}
\toprule
Symbol & Meaning & Value\\
\midrule
$P_{\rm LD}$ & LD optical power & 500 mW\\
$T_d$ & diffuser transmission & 0.9\\
$\Phalf$ & diffuser semi-angle & $20^{\circ}$ (desk) / $60^{\circ}$\\
$D_s$ & diffuser (source) diameter & 25 mm\\
RIN & laser intensity noise & $-120$ dB/Hz ($-110..-130$)\\
$A$ & photodiode area (150-$\mu$m PIN) & 7 mm$^{2}$ ($C_d{=}4.9$ pF)\\
$\mathcal{R}$ & responsivity (650 nm) & 0.45 A/W\\
$T_s$ & filter transmission & 0.85\\
$n$, $\Psi_c$ & concentrator index, FoV & 1.5; $15^{\circ}$/$60^{\circ}$\\
$B$ & receiver bandwidth $=$ symbol rate & 150 MHz\\
$R_f$, $C_T$ & TIA feedback R, input C & 5 k$\Omega$, 6.5 pF\\
GBW & required amplifier GBW \eqref{eq:tia} & 4.6 GHz\\
$i_n$, $v_n$ & TIA input noise densities & 3 pA, 1 nV$/\sqrt{\rm Hz}$\\
$I_{bg}$ & background photocurrent & 20 (desk) / 100 $\mu$A\\
FEC & HD tier / SD tier thresholds & $3.8{\times}10^{-3}$ / $2{\times}10^{-2}$\\
\bottomrule
\end{tabular}
\end{table}

\section{Modulation and FEC}\label{sec:modulation}

\subsection{Unipolar M-PAM}

The PHY uses baseband unipolar $M$-PAM at symbol rate $R_s=B$. This
choice is deliberate: intensity levels are directly the signal (no
electrical subcarrier), so there is no double-sideband bandwidth
penalty and no separately-accounted DC bias---the unipolarity cost is
inherent in the average-optical-power constraint of the constellation.
With levels $0,\dots,(M{-}1)\,\delta$, mean fixed at
$\mathcal{R}P_{r}$, and electrical SNR
$\gamma=(\mathcal{R}P_r)^2/\sigma^2$, Gray mapping gives
\begin{equation}
P_b \approx \frac{2(M-1)}{M\log_2 M}\,
Q\!\left(\frac{\sqrt{\gamma}}{M-1}\right),
\label{eq:pam}
\end{equation}
which reduces to $Q(\sqrt\gamma)$ for OOK ($M{=}2$). The $(M-1)$
penalty is steep---each doubling of $M$ costs $\sim$6--7~dB---which is
the honest price of IM/DD unipolarity; comparisons with DCO-OFDM
bit-loading, which can trade this differently at the cost of bias and
clipping analysis~\cite{islim2016modulation,niaz2017total}, are future
work on top of the released framework.

\subsection{Specified FEC Tiers}

\begin{figure}[!t]
\centering
\includegraphics[width=\columnwidth]{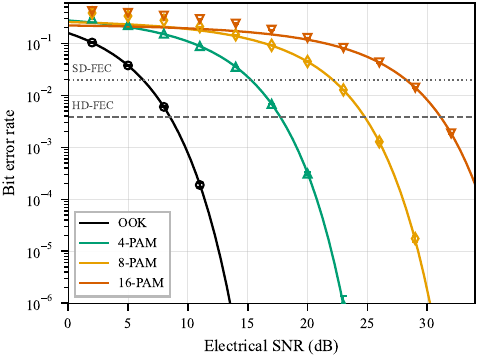}
\caption{Unipolar $M$-PAM BER versus electrical SNR: analysis (lines)
and Monte-Carlo counts with 95\% confidence intervals (markers;
$\ge$10 error events per point, up to $2\times10^{7}$ bits). Dashed/
dotted lines: HD and SD FEC thresholds.}
\label{fig:bersnr}
\end{figure}

Two concrete tiers, with the code family, overhead, and threshold
stated: (i) \emph{HD tier}: G.975.1-class concatenated hard-decision
FEC, 7\% overhead, pre-FEC threshold $3.8\times10^{-3}$ for
$<10^{-15}$ output~\cite{itu_g9751}; (ii) \emph{SD tier}: LDPC-class
soft-decision FEC, 20\% overhead, threshold
$2\times10^{-2}$---the pairing standard in optical transport. Net rate
of modulation $M$ under tier $t$: $R_{\rm net}=B\log_2M\,(1-\rho_t)$
when $P_b\le\mathrm{thr}_t$. Every analytic curve is checked by
Monte-Carlo simulation with error-count-based 95\% confidence
intervals (Fig.~\ref{fig:bersnr}); points are plotted only where at
least ten error events were counted.

\section{Eye Safety of the Diffused Source}\label{sec:safety}

\begin{figure}[!t]
\centering
\includegraphics[width=\columnwidth]{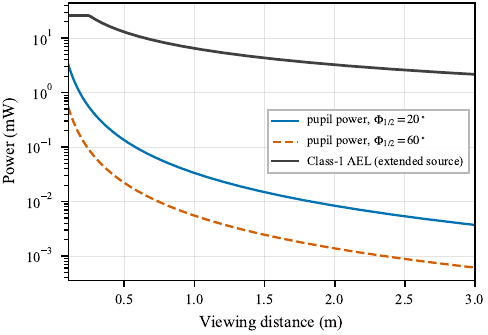}
\caption{Extended-source eye-safety assessment: 7-mm-pupil collected
power versus the Class-1 extended-source AEL as a function of viewing
distance, for both diffuser angles at $P_t=450$~mW. Minimum margin
7.8$\times$ (at 10 cm, $20^\circ$).}
\label{fig:safety}
\end{figure}

The diffuser converts the LD into an extended source of diameter
$D_s=25$~mm. Following the extended-source provisions of
IEC~60825-1~\cite{iec60825}, we compare the power collectable by a 7-mm
pupil at viewing distance $r$,
$P_{\rm pup}=(m{+}1)P_t A_{\rm pup}/2\pi r^2$, against the Class-1
accessible-emission limit $0.39\,\mathrm{mW}\times C_6$ with
$C_6=\alpha/1.5$~mrad, $\alpha=\mathrm{clamp}(D_s/r,\,1.5,\,100)$~mrad.
Fig.~\ref{fig:safety} shows the result for both diffuser angles: the
pupil power tracks a factor $\ge$7.8 \emph{below} the limit at all
distances $\ge$10~cm---the $20^\circ$ diffuser's worst margin occurs
nearest the source, where the $\alpha$ cap binds. This is an
engineering assessment, not certification (aided viewing/Class-1M
conditions remain for a product process), but it establishes that the
450-mW radiated power used throughout the link budget is compatible
with Class-1-style naked-eye exposure, closing the safety-performance
loop quantitatively. One single-fault condition must be stated
explicitly: a cracked or detached diffuser leaves a collimated 500-mW
visible emitter---Class~4. Any physical realization therefore
requires a diffuser-integrity interlock that removes drive current on
fault, and every ``eye-safer'' statement in this paper is conditioned
on that interlock. Two further scope notes follow the classification
literature. First, a laser-illuminated diffuse source may fall to be
assessed as a lamp rather than as a laser product, under
IEC~62471~\cite{iec62471_2006} and its visible-source
part~\cite{iec62471_7}, and the boundary between the two regimes for
laser-illuminated luminaires is itself the subject of dedicated
analysis~\cite{schulmeister2017illum,schulmeister2016rg2}; a product
realization must determine which applies. Second, eye-safe operation
of laser white lighting at multi-Gb/s is not hypothetical---it has
been demonstrated with tricolour R/G/B diodes~\cite{wu2017eyesafe}.

\section{Cross-Verified Simulation Framework}\label{sec:simulator}

\begin{figure}[!t]
\centering
\begin{tikzpicture}[
  node distance=2.2mm and 3mm,
  cls/.style={draw, rounded corners=1pt, minimum height=5.5mm,
              minimum width=21mm, align=center, font=\scriptsize},
  grp/.style={draw, dashed, rounded corners=2pt, inner sep=2mm},
  arr/.style={-{Stealth[length=1.5mm]}, semithick}]
\node[cls, fill=green!8] (app) {Applications, IPv4/UDP/TCP,\\ FlowMonitor (stock ns-3)};
\node[cls, fill=orange!10, below=of app] (dev) {\texttt{LifiNetDevice}};
\node[cls, fill=orange!10, below=of dev] (mac) {\texttt{LifiMac}\\ stop-and-wait ARQ (lower bound)};
\node[cls, fill=orange!10, below=of mac] (phy) {\texttt{LifiPhy}\\ full noise \eqref{eq:noise}, PAM BER};
\node[cls, fill=blue!8, below=of phy] (chn) {\texttt{LifiChannel} $+$\\ \texttt{LifiOpticalChannelModel}};
\node[cls, fill=gray!10, right=6mm of chn.north east, anchor=north west,
      minimum width=17mm] (mob) {\texttt{Mobility}\\ \texttt{Model}s};
\draw[arr] (app) -- (dev);
\draw[arr] (dev) -- (mac);
\draw[arr] (mac) -- (phy);
\draw[arr] (phy) -- (chn);
\draw[arr] (mob.west) -- ++(-3mm,0) |- (chn.east);
\node[grp, fit=(dev)(mac)(phy)(chn), label={[font=\scriptsize\itshape]right:\texttt{lifi} module}] {};
\end{tikzpicture}
\caption{ns-3 \texttt{lifi} module (v2). The MAC is an analyzable
lower bound, not a standards model; Section~\ref{sec:simulator}
relates it to IEEE 802.11bb and ITU-T G.9991.}
\label{fig:ns3arch}
\end{figure}
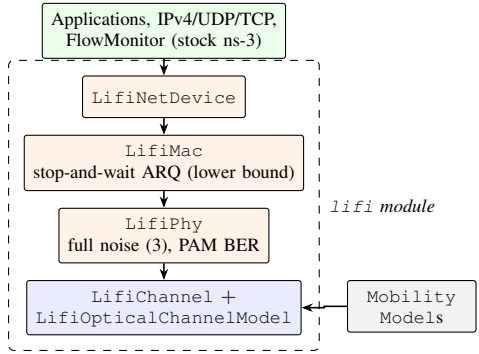

The framework comprises the ns-3 module of Fig.~\ref{fig:ns3arch} and
an independent Python engine implementing
\eqref{eq:H0}--\eqref{eq:pam}, the patch-integration multipath, and
the safety assessment. Automated tests pin the two implementations
together at 0.1\% on reference values (channel gain, SNR, thresholds).
We state its epistemic status precisely, in the established sense of
the verification-and-validation
literature~\cite{oberkampf2010vv,sargent2013vv,roache1998vv,
aiaa1998vv,asme2019vv}: agreement between the implementations, and
between analysis and Monte-Carlo counting, is
\emph{verification}---evidence that the equations are solved
correctly, and protection against transcription and coding error.
It is not \emph{validation}, which concerns whether the equations
describe the physical system
(Lambertian diffusion, linear LD response above threshold, the RIN
figure, speckle averaging), which only measurements can provide
(Section~\ref{sec:scope}).

\emph{Duplexing and MAC.} For the fixed desk terminals that carry all
network results, the return link is specified as a \emph{symmetric}
diffused-LD transceiver on a separated wavelength (e.g., 850~nm): same
radiated power class, same up-facing-luminaire geometry, hence the
same link budget and the same extended-source safety assessment as the
downlink---the prototype itself was symmetric in exactly this sense.
Acknowledgements therefore travel at the downlink line rate, and the
saturation formula below uses that specification. (Wide-FoV mobile
terminals would need a reduced-rate infrared uplink whose budget and
terminal-side safety are future work; the network results are stated
for fixed-terminal cells only.) The stop-and-wait MAC with bounded
retransmission is chosen as a fully analyzable \emph{lower bound}
whose saturation goodput
$t_{\rm pkt}/(t_{\rm pkt}{+}t_{\rm ack}{+}2\tau_p)$ any practical
aggregating protocol (802.11bb block-ack~\cite{ieee80211bb}, G.9991
data-link layer~\cite{itu_g9991}) will exceed. The module is a
physical-layer research vehicle beneath standard stacks, not a
standards implementation; mapping the PHY abstractions onto an
802.11bb MAC inside ns-3's WiFi model is planned and the module's
interfaces were designed for it.

\section{Results}\label{sec:results}

\subsection{SNR: the Cost of Realism}

\begin{figure}[!t]
\centering
\includegraphics[width=\columnwidth]{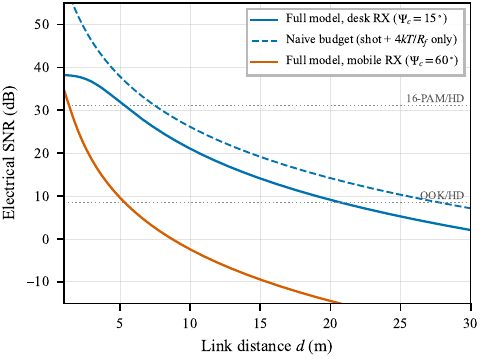}
\caption{Electrical SNR versus distance ($\Phalf=20^\circ$): full
budget \eqref{eq:noise} versus the naive shot$+4kT/R_f$ budget at
identical parameters (5.1-dB gap), and the wide-FoV mobile profile.
Dotted: OOK and 16-PAM HD thresholds.}
\label{fig:snrdist}
\end{figure}

Fig.~\ref{fig:snrdist} quantifies the central methodological point: at
14~m the full budget yields 15.3~dB where the naive budget claims
20.3~dB. Five decibels is the difference between 4-PAM and OOK at
range---rate factors, not rounding. Compared with the earlier
\emph{unconstrained} receiver (which additionally assumed an
unrealizable 250-MHz/1-cm$^2$/10-k$\Omega$ front end), the honest
accounting lowers projected net rate at 14~m from 930 to 140--240~Mb/s:
a factor 3.9--6.6 inflation that the realizability discipline of
Section~\ref{sec:receiver} exists to prevent.

\subsection{Error Rate, Reach, and Rate}

\begin{figure}[!t]
\centering
\includegraphics[width=\columnwidth]{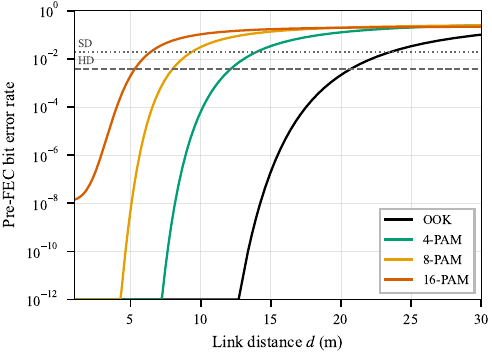}
\caption{Pre-FEC BER versus distance per modulation (desk profile).
HD-tier reaches: 5.3~m (16-PAM), 8.0~m (8-PAM), 12.1~m (4-PAM),
20.7~m (OOK); SD tier extends these by 14--22\% (23.6, 14.0, 9.4, and 6.5~m).}
\label{fig:berdist}
\end{figure}

\begin{figure}[!t]
\centering
\includegraphics[width=\columnwidth]{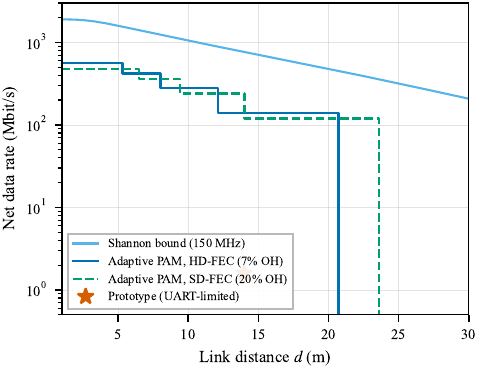}
\caption{Model-predicted net rate versus distance under both FEC
tiers, with the Shannon bound and the prototype's measured
UART-limited point. All staircase values are predictions per
Section~\ref{sec:scope}.}
\label{fig:ratedist}
\end{figure}

Figs.~\ref{fig:berdist} and \ref{fig:ratedist} give the revised
staircases. At the prototype's demonstrated 14~m: OOK/HD 140~Mb/s
with 6.7~dB of margin, or 4-PAM/SD 240~Mb/s---the latter at zero
margin, since the 4-PAM/SD reach is exactly 14.0~m (a 1-dB
implementation allowance retracts it to 13.2~m); the robust 14-m
operating point is therefore the OOK/HD figure. At 5~m: 16-PAM 480
(SD) to 558 (HD)~Mb/s. OOK persists to 20.7~m (HD) / 23.6~m (SD). The Shannon gap of
2.5--4~b/s/Hz locates the headroom accessible to DCO-OFDM bit-loading
and stronger codes; unlike the earlier version we do not treat that
observation as a contribution, merely as the map's edge.

\subsection{RIN Sensitivity}

\begin{figure}[!t]
\centering
\includegraphics[width=\columnwidth]{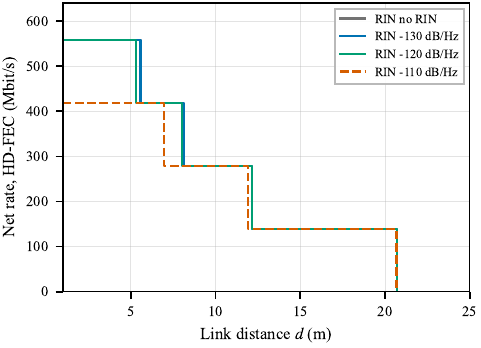}
\caption{HD-tier net rate versus distance as laser RIN varies from
$-130$ to $-110$~dB/Hz. At $-110$~dB/Hz the RIN ceiling (28.2~dB)
eliminates 16-PAM entirely; measuring the actual device RIN is
therefore a gating step for any high-order deployment.}
\label{fig:rin}
\end{figure}

Fig.~\ref{fig:rin} propagates the least-known parameter. The
conclusions that survive the full RIN range: OOK/4-PAM reaches move by
$<$0.5~m; the 16-PAM tier is fragile (absent at $-110$~dB/Hz). We
flag the highest-order predictions accordingly rather than assert
them.

\subsection{Coverage}

\begin{figure*}[!t]
\centering
\includegraphics[width=\textwidth]{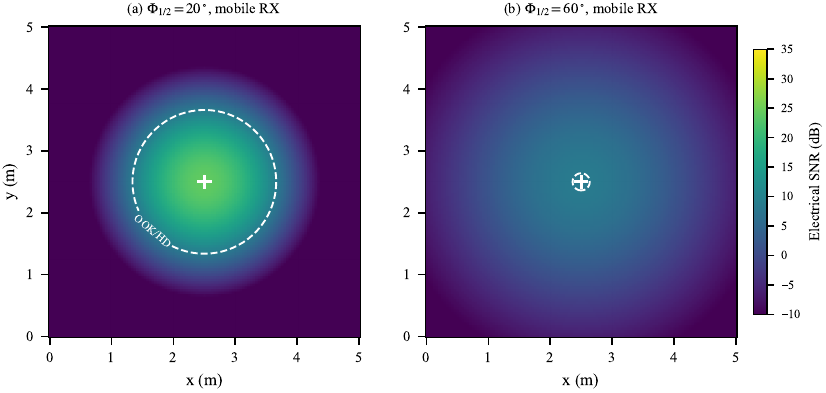}
\caption{Desk-plane SNR maps for the mobile receiver profile
($\Psi_c=60^\circ$): (a) $\Phalf=20^\circ$ concentrates a small
high-rate cell under the luminaire; (b) $\Phalf=60^\circ$ trades peak
SNR for wall-to-wall OOK coverage. The dashed contour is the OOK/HD
boundary.}
\label{fig:coverage}
\end{figure*}

Fig.~\ref{fig:coverage} maps the room with the mobile profile. The
wide-FoV terminal cannot close a 14-m boresight link
(Fig.~\ref{fig:snrdist}); its regime is the under-luminaire cell,
where the $20^\circ$ source supports high-order PAM over a
$\sim$1-m-radius spot and the $60^\circ$ source supports OOK
essentially wall to wall. Multi-luminaire tiling and handover atop
these cells are exactly the studies the ns-3 module exists to host.

\subsection{Fidelity and Network Behavior}

\begin{figure}[!t]
\centering
\includegraphics[width=\columnwidth]{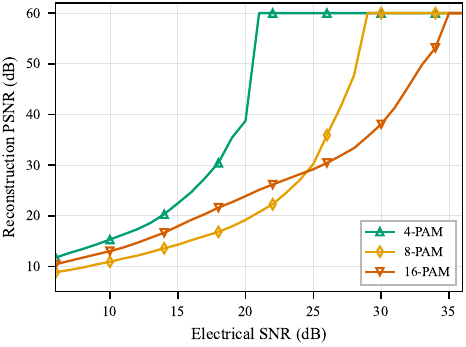}
\caption{Bit-exact reconstruction PSNR of the prototype's
$69\times71$ test image versus SNR (uncoded PAM chain).}
\label{fig:psnr}
\end{figure}

\begin{figure*}[!t]
\centering
\includegraphics[width=\textwidth]{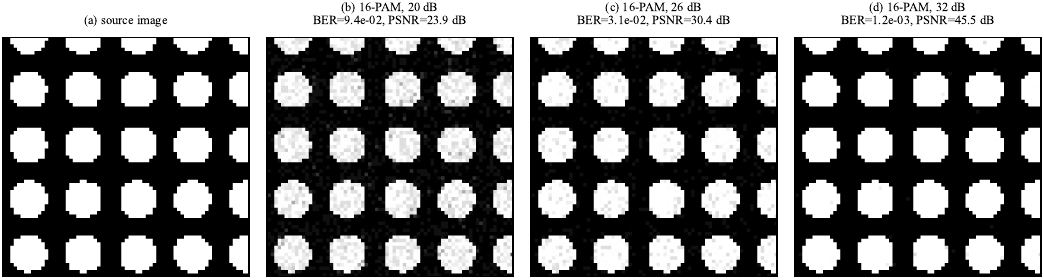}
\caption{Uncoded 16-PAM image transfer at three SNRs (b--d); with
either FEC tier above threshold, delivery is bit-exact, matching the
prototype's observed error-free transfers at its (deeply
threshold-clearing) UART rates.}
\label{fig:imagepanel}
\end{figure*}

\begin{figure}[!t]
\centering
\includegraphics[width=\columnwidth]{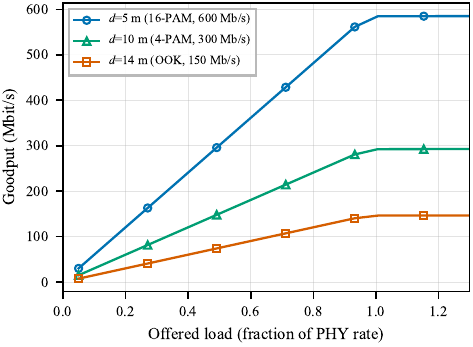}
\caption{Goodput versus offered load, stop-and-wait lower bound,
full-duplex ACK wavelength, 1500-B packets, post-FEC residual BER
$10^{-9}$.}
\label{fig:load}
\end{figure}

\begin{figure}[!t]
\centering
\includegraphics[width=\columnwidth]{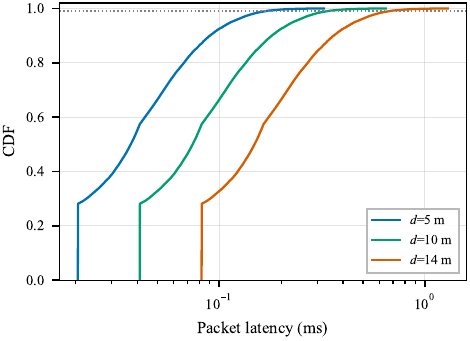}
\caption{Packet-latency CDFs at 70\% load. 99th-percentile: 0.17~ms
(5~m), 0.33~ms (10~m), 0.67~ms (14~m).}
\label{fig:latency}
\end{figure}

The image experiment of the prototype is reproduced bit-exactly
through the PAM chain (Figs.~\ref{fig:psnr}--\ref{fig:imagepanel}).
Network-level curves (Figs.~\ref{fig:load}--\ref{fig:latency}) carry
the stated MAC lower bound over the revised PHY rates: saturation at
93--96\% of line rate and sub-millisecond 99th-percentile latency at
70\% load. These inherit the PHY model's epistemic status; they
demonstrate the framework and bound the protocol overhead, and they
will sharpen (upward in throughput) under 802.11bb-style aggregation.

\subsection{A Fair Emitter Comparison}

At matched radiated power (450~mW, eye-safe for both emitter types),
matched receiver, and honest bandwidths---15~MHz for a
phosphor-white LED lamp versus the LD's (receiver-limited) 150~MHz---
the model gives 14-m HD rates of $\sim$28~Mb/s (LED, 4-PAM at 15~MHz)
versus 140~Mb/s (LD): a factor $\sim$5 attributable to bandwidth, not
a contrived power gap. We note explicitly that engineered LED systems
with WDM and adaptive loading exceed both figures at short
range~\cite{bian2019lifi}; the LD case rests on retaining its
bandwidth advantage at range and modest complexity, not on beating
LED flagships.

\section{What Is Measured, What Is Predicted, and What Is Next}
\label{sec:scope}

\begin{table}[!t]
\caption{Claims Ledger: Evidence Class of Every Headline Number}
\label{tab:claims}
\centering
\scriptsize
\begin{tabular}{p{0.44\columnwidth}p{0.20\columnwidth}p{0.22\columnwidth}}
\toprule
Claim & Class & Basis\\
\midrule
End-to-end data/voice/image over 14 m & measured &
prototype~\cite{ahmad2019thesis}\\
Goodput ceiling 1.59 Mb/s (CH340 path) & measured/analytic &
UART framing (Fig.~\ref{fig:uart})\\
Zero observed errors at 14 m, UART rates & measured; consistent &
model predicts BER $<10^{-12}$ there\\
SNR 15.3 dB at 14 m, $B{=}150$ MHz & predicted &
full budget \eqref{eq:noise}\\
140--240 Mb/s net at 14 m & predicted &
PAM + FEC tiers\\
480--558 Mb/s net at 5 m & predicted (RIN-fragile) &
Fig.~\ref{fig:rin}\\
Multipath $K\ge4.2$ dB, $\tau_{\rm rms}\le0.13$ ns & computed &
first-bounce integration\\
Class-1-style margin $\ge$7.8$\times$ & computed &
Section~\ref{sec:safety}\\
\bottomrule
\end{tabular}
\end{table}

\begin{figure}[!t]
\centering
\includegraphics[width=\columnwidth]{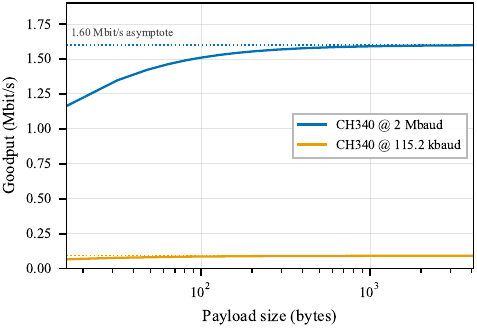}
\caption{The prototype's measured regime: CH340 UART goodput ceiling
(1.59~Mb/s at 1500-B payloads). Timing-based estimates exceeding this
(the thesis's 127~Mb/s~\cite{ahmad2019thesis}) are processing-side
artifacts, not link throughput.}
\label{fig:uart}
\end{figure}

Table~\ref{tab:claims} is the paper's contract with the reader: every
headline number labeled by evidence class. The prototype validates
feasibility and geometry; it cannot validate the wideband model, whose
observation of zero errors at UART rates is \emph{consistent with} but
not \emph{discriminating between} parameterizations---a limitation we
state rather than blur. The experimental path this framework
specifies for its own confirmation, in order of information gained per
effort: (i) measured end-to-end frequency response of an LD-driver/
PIN-TIA chain built to Table~\ref{tab:params}; (ii) measured RIN of
the deployed LD class; (iii) BER-versus-distance at 50--150~MBd OOK/
4-PAM, off the error floor; (iv) received-plane scans versus the
Lambertian fit of the actual diffuser, including speckle statistics.
Items (i)--(iii) require only commodity instruments (VNA-lite,
BER tester at 150~MBd); none require the multi-GHz apparatus of
record-chasing demonstrations.

Model scope that remains open: LD nonlinearity and clipping for
high-order PAM; speckle and modal noise statistics of static
holographic diffusers under coherent illumination; higher-order
reflection bounces; the mobile-terminal infrared uplink; and an
802.11bb MAC mapping.

\section{Conclusion}

This paper asked what a diffused-beam LD LiFi link supports when every
optimistic assumption is replaced by an enforceable constraint: a
buildable TIA, a complete noise budget with RIN, unipolar PAM without
spectral sleight of hand, specified FEC, computed multipath, and a
quantitative safety margin at the actual operating power. The answer
---140--240~Mb/s at 14~m, half a gigabit at 5~m, OOK to 21~m, with the
top tier explicitly RIN-fragile---is smaller than idealized budgets
suggest by factors of 3.9--6.6, and that correction, made auditable
and reproducible through an open cross-verified ns-3/Python framework,
is the contribution. The framework's claims ledger separates its one
measured anchor from its predictions, and its specified confirmation
experiments are deliberately modest. We believe this accounting
discipline---realizability frontiers, complete budgets, evidence-class
labeling---transfers directly to the broader indoor-optical-wireless
modeling literature, where the gap this paper closes for one system
remains open for many.

\bibliographystyle{IEEEtran}
\bibliography{refs}

@article{gao2018real,
  author  = {Gao, Y.-L. and Wu, Z.-Y. and Wang, Z.-K. and Wang, J.},
  title   = {A 1.34-{Gb/s} Real-Time {Li-Fi} Transceiver With {DFT}-Spread-Based {PAPR} Mitigation},
  journal = {IEEE Photonics Technology Letters},
  volume  = {30},
  number  = {16},
  pages   = {1447--1450},
  year    = {2018}
}

@article{islim2016modulation,
  author  = {Islim, M. S. and Haas, H.},
  title   = {Modulation Techniques for {Li-Fi}},
  journal = {ZTE Communications},
  volume  = {14},
  number  = {2},
  pages   = {29--40},
  year    = {2016}
}

@article{chi2015gan,
  author  = {Chi, Y.-C. and Hsieh, D.-H. and Tsai, C.-T. and Chen, H.-Y. and Kuo, H.-C. and Lin, G.-R.},
  title   = {450-nm {GaN} Laser Diode Enables High-Speed Visible Light Communication with 9-{Gbps} {QAM-OFDM}},
  journal = {Optics Express},
  volume  = {23},
  number  = {10},
  pages   = {13051--13059},
  year    = {2015}
}

@article{niaz2017total,
  author  = {Niaz, M. T. and Imdad, F. and Kim, S. and Kim, H. S.},
  title   = {Total Least-Square-Based Receiver for Asymmetrically Clipped Optical-{OFDM} Visible Light Communication System},
  journal = {IET Optoelectronics},
  volume  = {11},
  number  = {4},
  pages   = {129--133},
  year    = {2017}
}

@article{wu2018white,
  author  = {Wu, T.-C. and Chi, Y.-C. and Wang, H.-Y. and Tsai, C.-T. and Huang, Y.-F. and Lin, G.-R.},
  title   = {White-Lighting Communication With a {Lu$_3$Al$_5$O$_{12}$:Ce$^{3+}$/CaAlSiN$_3$:Eu$^{2+}$} Glass Covered 450-nm {InGaN} Laser Diode},
  journal = {Journal of Lightwave Technology},
  volume  = {36},
  number  = {9},
  pages   = {1634--1643},
  year    = {2018}
}

@article{chi2017violet,
  author  = {Chi, Y.-C. and Huang, Y.-F. and Wu, T.-C. and Tsai, C.-T. and Chen, L.-Y. and Kuo, H.-C. and Lin, G.-R.},
  title   = {Violet Laser Diode Enables Lighting Communication},
  journal = {Scientific Reports},
  volume  = {7},
  number  = {1},
  pages   = {10469},
  year    = {2017}
}

@article{kahn1997wireless,
  author  = {Kahn, J. M. and Barry, J. R.},
  title   = {Wireless Infrared Communications},
  journal = {Proceedings of the IEEE},
  volume  = {85},
  number  = {2},
  pages   = {265--298},
  year    = {1997}
}

@article{komine2004fundamental,
  author  = {Komine, T. and Nakagawa, M.},
  title   = {Fundamental Analysis for Visible-Light Communication System Using {LED} Lights},
  journal = {IEEE Transactions on Consumer Electronics},
  volume  = {50},
  number  = {1},
  pages   = {100--107},
  year    = {2004}
}

@article{haas2016what,
  author  = {Haas, H. and Yin, L. and Wang, Y. and Chen, C.},
  title   = {What is {LiFi}?},
  journal = {Journal of Lightwave Technology},
  volume  = {34},
  number  = {6},
  pages   = {1533--1544},
  year    = {2016}
}

@misc{iec60825,
  author       = {{IEC}},
  title        = {Safety of Laser Products---Part 1: Equipment Classification and Requirements},
  howpublished = {IEC 60825-1:2014},
  year         = {2014}
}

@article{carruthers1997modeling,
  author  = {Carruthers, J. B. and Kahn, J. M.},
  title   = {Modeling of Nondirected Wireless Infrared Channels},
  journal = {IEEE Transactions on Communications},
  volume  = {45},
  number  = {10},
  pages   = {1260--1268},
  year    = {1997}
}

@article{barry1993simulation,
  author  = {Barry, J. R. and Kahn, J. M. and Krause, W. J. and Lee, E. A. and Messerschmitt, D. G.},
  title   = {Simulation of Multipath Impulse Response for Indoor Wireless Optical Channels},
  journal = {IEEE Journal on Selected Areas in Communications},
  volume  = {11},
  number  = {3},
  pages   = {367--379},
  year    = {1993}
}

@book{sackinger2005broadband,
  author    = {S{\"a}ckinger, E.},
  title     = {Broadband Circuits for Optical Fiber Communication},
  publisher = {Wiley},
  year      = {2005}
}

@book{petermann1988laser,
  author    = {Petermann, K.},
  title     = {Laser Diode Modulation and Noise},
  publisher = {Kluwer Academic},
  year      = {1988}
}

@misc{itu_g9751,
  author       = {{ITU-T}},
  title        = {Forward Error Correction for High Bit-Rate {DWDM} Submarine Systems},
  howpublished = {Recommendation G.975.1},
  year         = {2004}
}

@misc{itu_g9991,
  author       = {{ITU-T}},
  title        = {High-Speed Indoor Visible Light Communication Transceiver---System Architecture, Physical Layer and Data Link Layer Specification},
  howpublished = {Recommendation G.9991},
  year         = {2019}
}

@misc{ieee80211bb,
  author       = {{IEEE}},
  title        = {{IEEE} Standard for Information Technology---Wireless {LAN} {MAC} and {PHY} Specifications---Amendment 7: Light Communications},
  howpublished = {IEEE Std 802.11bb-2023},
  year         = {2023}
}

@article{haas2020introduction,
  author  = {Haas, H. and Yin, L. and Chen, C. and Videv, S. and Parol, D. and Poves, E. and Alshaer, H. and Islim, M. S.},
  title   = {Introduction to Indoor Networking Concepts and Challenges in {LiFi}},
  journal = {Journal of Optical Communications and Networking},
  volume  = {12},
  number  = {2},
  pages   = {A190--A203},
  year    = {2020}
}

@article{bian2019lifi,
  author  = {Bian, Rui and Tavakkolnia, Iman and Haas, Harald},
  title   = {15.73 {G}b/s Visible Light Communication With Off-the-Shelf {LED}s},
  journal = {Journal of Lightwave Technology},
  volume  = {37},
  number  = {10},
  pages   = {2418--2424},
  year    = {2019},
  doi     = {10.1109/JLT.2019.2906464}
}

@article{koonen2018indoor,
  author  = {Koonen, T.},
  title   = {Indoor Optical Wireless Systems: Technology, Trends, and Applications},
  journal = {Journal of Lightwave Technology},
  volume  = {36},
  number  = {8},
  pages   = {1459--1467},
  year    = {2018}
}

@mastersthesis{ahmad2019thesis,
  author = {Ahmad, H. and Gillani, S. M. T. and Omer, T.},
  title  = {High Speed {LiFi} Transceiver},
  school = {Department of Electrical Engineering, Bahria University, Islamabad},
  type   = {{BS} thesis},
  year   = {2019}
}

@inproceedings{ahmad2019futuristic,
  author    = {Ahmad, H. and Gillani, S. M. T. and Omer, T. and Hassan, T. and Aslam, S. and Ali, S. U.},
  title     = {Futuristic Short Range Optical Communication: A Survey},
  booktitle = {Proc. International Conference on Information Science and Communication Technology (ICISCT)},
  year      = {2019}
}

@article{chen2024jlt100g,
  author  = {Chen, Cheng and Das, Sujan and Videv, Stefan and Sparks, Andrew and Babadi, Sanaz and Krishnamoorthy, Anil and Lee, Chang and Grieder, Douglas and Hartnett, Kathleen P. and Rudy, Paul and Raring, James W. and Najafi, Marzieh and Papanikolaou, Vasilis K. and Schober, Robert and Haas, Harald},
  title   = {100 {G}bps Indoor Access and 4.8 {G}bps Outdoor Point-to-Point {L}i{F}i Transmission Systems Using Laser-Based Light Sources},
  journal = {Journal of Lightwave Technology},
  volume  = {42},
  number  = {12},
  pages   = {4146--4157},
  year    = {2024},
  doi     = {10.1109/JLT.2024.3400192}
}

@article{wang2024gan5ghz,
  author  = {Wang, Junfei and Hu, Junhui and Guan, Chenyang and Hou, Yuqi and Xu, Zengyi and Sun, Leihao and Wang, Yue and Zhou, Yingjun and Ooi, Boon S. and Shi, Jianyang and Li, Ziwei and Zhang, Junwen and Chi, Nan and Yu, Shaohua and Shen, Chao},
  title   = {High-speed {G}a{N}-based laser diode with modulation bandwidth exceeding 5 {GHz} for 20 {G}bps visible light communication},
  journal = {Photonics Research},
  volume  = {12},
  number  = {6},
  pages   = {1186--1193},
  year    = {2024},
  doi     = {10.1364/PRJ.516829}
}

@article{jia2025blueld,
  author  = {Jia, Haolin and Hu, Junhui and Xu, Zengyi and Gu, Zhenqian and Yang, Zhen and Li, Zengxin and Zhou, Yingjun and Shi, Jianyang and Li, Ziwei and Zhang, Junwen and Chi, Nan and Shen, Chao},
  title   = {High-Speed Blue Laser Diodes with {I}n{G}a{N} Quantum Barrier for Beyond 36 {G}bps Visible Light Communications},
  journal = {Laser {\&} Photonics Reviews},
  volume  = {19},
  number  = {11},
  pages   = {2401751},
  year    = {2025},
  doi     = {10.1002/lpor.202401751}
}

@article{hu2026minild,
  author  = {Hu, Junhui and Gu, Zhenqian and Jia, Haolin and Yang, Zhen and Li, Zengxin and Wu, Jiabin and Sun, Leihao and Sun, Aolong and Huang, Ouhan and Xia, Changsheng and Ooi, Boon S. and Shi, Jianyang and Li, Ziwei and Zhang, Junwen and Yu, Shaohua and Chi, Nan and Shen, Chao},
  title   = {{I}n{G}a{N}-based high-speed mini laser diode surpasses {PAM}-4 visible light links by over 30 {G}bps},
  journal = {Science China Information Sciences},
  volume  = {69},
  number  = {3},
  pages   = {132405},
  year    = {2026},
  doi     = {10.1007/s11432-025-4581-8}
}

@article{chi2025wdm500,
  author  = {Chi, Nan and Niu, Wenqing and Zhou, Yingjun and Wang, Junfei and Chen, Hui and He, Zhixue and Li, Jiali and Xu, Zengyi and Lin, Xianhao and Luo, Zhiteng and Lu, Zhilan and Zhang, Junwen and Shen, Chao and Li, Ziwei and Shi, Jianyang and Yu, Shaohua},
  title   = {Enabling Technologies to Achieve Beyond 500 {G}bps Optical Intra-Connects Based on {WDM} Visible Light Laser Communication},
  journal = {Journal of Lightwave Technology},
  volume  = {43},
  number  = {4},
  pages   = {1843--1854},
  year    = {2025},
  doi     = {10.1109/JLT.2024.3486062}
}

@article{zhou2025col600,
  author  = {Zhou, Yingjun and Lin, Xianhao and Xu, Zengyi and Lu, Zhilan and Huang, Yifan and Luo, Zhiteng and Zhang, Hao and Sun, Leihao and Ye, Jiabin and Chen, Hui and He, Zhixue and Li, Jiali and Zhang, Junwen and Shen, Chao and Yu, Shaohua and Chi, Nan},
  title   = {Beyond 600 {G}bps optical interconnect utilizing wavelength division multiplexed visible light laser communication [{Invited}]},
  journal = {Chinese Optics Letters},
  volume  = {23},
  number  = {5},
  pages   = {050002},
  year    = {2025},
  doi     = {10.3788/COL202523.050002}
}

@article{luo2024rainbow,
  author  = {Luo, Zhiteng and Lin, Xianhao and Lu, Zhilan and Shi, Jianyang and Shen, Chao and Li, Ziwei and Zhang, Junwen and He, Zhixue and Chen, Hui and Yu, Shaohua and Chi, Nan},
  title   = {113{G}bps rainbow visible light laser communication system based on 10{$\lambda$} laser {WDM} emitting module in fiber-free space-fiber link},
  journal = {Optics Express},
  volume  = {32},
  number  = {2},
  pages   = {2561--2573},
  year    = {2024},
  doi     = {10.1364/OE.508065}
}

@article{yaseen2025rin,
  author  = {Yaseen, Maysa and Elamassie, Mohammed and Ikki, Salama S. and Uysal, Murat},
  title   = {Signal-Dependent Shot and Relative Intensity Noise in Channel Estimation of Laser Diode-Based Indoor {VLC} Systems},
  journal = {IEEE Transactions on Communications},
  volume  = {73},
  number  = {1},
  pages   = {498--509},
  year    = {2025},
  doi     = {10.1109/TCOMM.2024.3420736}
}

@article{lazzaro2026speckle,
  author  = {Lazzaro, Joseph P. and Wang, Xiaoxin and Carver, Charles J. and Shade, Nicholas R. and Schwartz, Hannah and Liu, Yang and Zhou, Xia and Liu, Jifeng and Fossum, Eric R.},
  title   = {Multi-{G}bps visible light communication with high-efficiency, low-speckle contrast white laser light},
  journal = {Optical Review},
  volume  = {33},
  number  = {1},
  pages   = {135--141},
  year    = {2026},
  doi     = {10.1007/s10043-025-01023-5}
}

@article{soltani2022safety,
  author  = {Soltani, Mohammad Dehghani and Sarbazi, Elham and Bamiedakis, Nikolaos and de Souza, Priscila and Kazemi, Hossein and Elmirghani, Jaafar M. H. and White, Ian H. and Penty, Richard V. and Haas, Harald and Safari, Majid},
  title   = {Safety Analysis for Laser-Based Optical Wireless Communications: A Tutorial},
  journal = {Proceedings of the IEEE},
  volume  = {110},
  number  = {8},
  pages   = {1045--1072},
  year    = {2022},
  doi     = {10.1109/JPROC.2022.3181968}
}

@article{hou2024tutorial,
  author  = {Hou, Yuqi and Wang, Yue and Li, Zengxin and Liu, Meixin and Yi, Shulan and Wang, Xiaoqian and Xia, Liang and Liu, Guangyi and Shi, Jianyang and Li, Ziwei and Zhang, Junwen and Chi, Nan and Ng, Tien Khee and Ooi, Boon S. and Shen, Chao},
  title   = {Tutorial on laser-based visible light communications [{Invited}]},
  journal = {Chinese Optics Letters},
  volume  = {22},
  number  = {9},
  pages   = {092502},
  year    = {2024},
  doi     = {10.3788/COL202422.092502}
}

@article{gu2026gan,
  author  = {Gu, Zhenqian and Yang, Zhen and Du, Fan and Fan, Ding and Xiao, Bo and Hu, Junhui and Jia, Haolin and Wang, Shuai and Pan, Yuan and Zhang, Junwen and Chi, Nan and Kafar, Anna and Perlin, Piotr and Shen, Chao},
  title   = {Advances of high-speed {G}a{N}-based laser diodes for emerging applications},
  journal = {Progress in Quantum Electronics},
  volume  = {108},
  pages   = {100633},
  year    = {2026},
  doi     = {10.1016/j.pquantelec.2026.100633}
}

@article{jin2025sdm,
  author  = {Jin, Zuxin and Ying, Hong and Zhang, Zhen and Lin, Runze and Gu, Erdan and Tian, Pengfei},
  title   = {Toward 20-{G}bps Single-Chip Space Division Multiplexing Visible Light Communication Based on a Low-Crosstalk Micro-{LED} Array},
  journal = {Journal of Lightwave Technology},
  volume  = {43},
  number  = {24},
  pages   = {10960--10967},
  year    = {2025},
  doi     = {10.1109/JLT.2025.3620260}
}

@article{qiu2022beyond25,
  author  = {Qiu, Pengjiang and Zhu, Shijie and Jin, Zuxin and Zhou, Xiaolin and Cui, Xugao and Tian, Pengfei},
  title   = {Beyond 25 {G}bps optical wireless communication using wavelength division multiplexed {LED}s and micro-{LED}s},
  journal = {Optics Letters},
  volume  = {47},
  number  = {2},
  pages   = {317--320},
  year    = {2022},
  doi     = {10.1364/OL.447540}
}

@article{tang2023rgbp,
  author  = {Tang, Lijuan and Wu, Yinmei and Cheng, Zerui and Teng, Dongdong and Liu, Lilin},
  title   = {Over 23.43 {G}bps visible light communication system based on 9 {V} integrated {RGBP} {LED} modules},
  journal = {Optics Communications},
  volume  = {534},
  pages   = {129317},
  year    = {2023},
  doi     = {10.1016/j.optcom.2023.129317}
}

@article{liu2021sdmwdm,
  author  = {Liu, Xin and Wei, Zixian and Li, Mutong and Wang, Lei and Liu, Zhongxu and Yu, Changyuan and Wang, Lai and Luo, Yi and Fu, H. Y.},
  title   = {Experimental investigation of 16.6 {G}bps {SDM}-{WDM} visible light communication based on a neural network receiver and tricolor mini-{LED}s},
  journal = {Optics Letters},
  volume  = {46},
  number  = {12},
  pages   = {2888--2891},
  year    = {2021},
  doi     = {10.1364/OL.428013}
}

@article{ai2025microled,
  author  = {Ai, Jiakang and Zhang, Zhen and Deng, Tianlin and Shen, Daqi and Xiao, Yang and Jin, Zuxin and Ren, Tianyang and Cui, Xugao and Tian, Pengfei},
  title   = {15.64 {G}bps High-Speed Visible Light Communication and Multiple-User Secure Communication Employing Red {G}a{N} Micro-{LED} Based Semi-Transparent Photodetector},
  journal = {Journal of Lightwave Technology},
  volume  = {43},
  number  = {15},
  pages   = {7058--7065},
  year    = {2025},
  doi     = {10.1109/JLT.2025.3568082}
}

@article{rao2024cplane,
  author  = {Rao, Zhiwei and Shan, Xinyi and Wang, Guobin and Jin, Zuxin and Lin, Runze and Cui, Xugao and Liu, Ran and Xu, Ke and Tian, Pengfei},
  title   = {10.5 {G}bps Visible Light Communication Systems Based on {C}-Plane Freestanding {G}a{N} Micro-{LED}},
  journal = {Journal of Lightwave Technology},
  volume  = {42},
  number  = {13},
  pages   = {4360--4364},
  year    = {2024},
  doi     = {10.1109/JLT.2024.3363729}
}

@article{jin2023tenm,
  author  = {Jin, Zuxin and Yan, Lintao and Zhu, Shijie and Cui, Xugao and Tian, Pengfei},
  title   = {10-{G}bps visible light communication in a 10-m free space based on violet series-biased micro-{LED} array and distance adaptive pre-equalization},
  journal = {Optics Letters},
  volume  = {48},
  number  = {8},
  pages   = {2026--2029},
  year    = {2023},
  doi     = {10.1364/OL.487747}
}

@article{jin2025green,
  author  = {Jin, Zuxin and Lin, Runze and Liu, Xuyang and Sun, Di and Liu, Bin and Tao, Tao and Xu, Feifan and Fang, Zhilai and Cui, Xugao and Tian, Pengfei},
  title   = {Green Micro-{LED} With a Bandwidth Exceeding 2 {GHz} for 9-{G}bps Visible Light Communication Based on {SNR} Gap-Dependent Bit and Power Loading},
  journal = {Journal of Lightwave Technology},
  volume  = {43},
  number  = {2},
  pages   = {472--480},
  year    = {2025},
  doi     = {10.1109/JLT.2024.3470242}
}

@article{wei2021ann,
  author  = {Wei, Zixian and Liu, Zhongxu and Liu, Xin and Wang, Lei and Wang, Lai and Yu, Changyuan and Fu, H. Y.},
  title   = {8.75 {G}bps visible light communication link using an artificial neural network equalizer and a single-pixel blue micro-{LED}},
  journal = {Optics Letters},
  volume  = {46},
  number  = {18},
  pages   = {4670--4673},
  year    = {2021},
  doi     = {10.1364/OL.437632}
}

@article{dong2026rc,
  author  = {Dong, Yixian and Huang, Yeping and Deng, Xiong and Li, Songsui and Zou, Xihua and Pan, Wei and Bienstman, Peter and Yan, Lianshan},
  title   = {{LED} Model-Driven Reservoir Computing for Signal Equalization in {OFDM}-{VLC} System},
  journal = {Journal of Lightwave Technology},
  volume  = {44},
  number  = {9},
  pages   = {3500--3512},
  year    = {2026},
  doi     = {10.1109/JLT.2026.3663361}
}

@article{yoshida2024oled,
  author  = {Yoshida, Kou and Chen, Cheng and Haas, Harald and Turnbull, Graham A. and Samuel, Ifor D. W.},
  title   = {{RGB}-Single-Chip {OLED}s for High-Speed Visible-Light Communication by Wavelength-Division Multiplexing},
  journal = {Advanced Science},
  volume  = {11},
  number  = {47},
  pages   = {2404576},
  year    = {2024},
  doi     = {10.1002/advs.202404576}
}

@article{sarbazi2024receivers,
  author  = {Sarbazi, Elham and Kazemi, Hossein and Crisp, Michael and El-Gorashi, Taisir and Elmirghani, Jaafar and Penty, Richard V. and White, Ian H. and Safari, Majid and Haas, Harald},
  title   = {Design and Optimization of High-Speed Receivers for 6{G} Optical Wireless Networks},
  journal = {IEEE Transactions on Communications},
  volume  = {72},
  number  = {2},
  pages   = {971--990},
  year    = {2024},
  doi     = {10.1109/TCOMM.2023.3328265}
}

@article{soltani2023imaging,
  author  = {Dehghani Soltani, Mohammad and Kazemi, Hossein and Sarbazi, Elham and El-Gorashi, Taisir E. H. and Elmirghani, Jaafar M. H. and Penty, Richard V. and White, Ian H. and Haas, Harald and Safari, Majid},
  title   = {High-Speed Imaging Receiver Design for 6{G} Optical Wireless Communications: A Rate-{FOV} Trade-Off},
  journal = {IEEE Transactions on Communications},
  volume  = {71},
  number  = {2},
  pages   = {1024--1043},
  year    = {2023},
  doi     = {10.1109/TCOMM.2022.3230954}
}

@article{portnoi2021lsc,
  author  = {Portnoi, Mark and Haigh, Paul Anthony and Macdonald, Thomas J. and Ambroz, Filip and Parkin, Ivan P. and Darwazeh, Izzat and Papakonstantinou, Ioannis},
  title   = {Bandwidth limits of luminescent solar concentrators as detectors in free-space optical communication systems},
  journal = {Light: Science {\&} Applications},
  volume  = {10},
  number  = {1},
  pages   = {3},
  year    = {2021},
  doi     = {10.1038/s41377-020-00444-y}
}

@article{meucci2024lsc,
  author  = {Meucci, Marco and Doria, Sandra and Umair, Ali Muhammad and Franchi, Daniele and Fattori, Marco and Di Donato, Mariangela and Picchi, Alberto and Pucci, Andrea and Calamante, Massimo and Catani, Jacopo},
  title   = {Efficient White-Light Visible Light Communication With Novel Optical Antennas Based on Luminescent Solar Concentrators},
  journal = {Journal of Lightwave Technology},
  volume  = {42},
  number  = {7},
  pages   = {2235--2244},
  year    = {2024},
  doi     = {10.1109/JLT.2023.3337040}
}

@article{ali2022sipm,
  author  = {Ali, Wajahat and Manousiadis, Pavlos P. and O'Brien, Dominic C. and Turnbull, Graham A. and Samuel, Ifor D. W. and Collins, Steve},
  title   = {A Gigabit {VLC} Receiver That Incorporates a Fluorescent Antenna and a {S}i{PM}},
  journal = {Journal of Lightwave Technology},
  volume  = {40},
  number  = {16},
  pages   = {5369--5375},
  year    = {2022},
  doi     = {10.1109/JLT.2021.3095398}
}

@article{matthews2023roadmap,
  author  = {Matthews, William and Collins, Steve},
  title   = {A Roadmap for Gigabit to Terabit Optical Wireless Communications Receivers},
  journal = {Sensors},
  volume  = {23},
  number  = {3},
  pages   = {1101},
  year    = {2023},
  doi     = {10.3390/s23031101}
}

@article{liu2023uvsipm,
  author  = {Liu, Fan and Farmer, Joshua and Schreier, Alexander and Faulkner, Grahame and Chun, Hyunchae and Matthews, William and Wang, Zhaoming and O'Brien, Dominic},
  title   = {Ultra-sensitive {UV} solar-blind optical wireless communications with an {S}i{PM}},
  journal = {Optics Letters},
  volume  = {48},
  number  = {20},
  pages   = {5387--5390},
  year    = {2023},
  doi     = {10.1364/OL.503235}
}

@article{bashir2024pdsize,
  author  = {Bashir, Muhammad Salman and Ahmed, Qasim Zeeshan and Alouini, Mohamed-Slim},
  title   = {Optimal Photodetector Size for High-Speed Free-Space Optics Receivers},
  journal = {IEEE Transactions on Wireless Communications},
  volume  = {23},
  number  = {11},
  pages   = {16390--16403},
  year    = {2024},
  doi     = {10.1109/TWC.2024.3440878}
}

@article{wang2026apdarray,
  author  = {Wang, Yijing and Xiong, Jian and Qi, Liuteng and Qi, Yuwen and Li, Jingzhou and Dong, Hongxing and Zhang, Long},
  title   = {Design and implementation of an avalanche photodiode array receiver for fluorescent-antenna-based visible light communication systems},
  journal = {Applied Optics},
  volume  = {65},
  number  = {11},
  pages   = {3483--3490},
  year    = {2026},
  doi     = {10.1364/AO.590014}
}

@article{meucci2025comparative,
  author  = {Meucci, Marco and Aresti, Mauro and Cossu, Giulio and Gilli, Lorenzo and Oliviero, Luca and Bartolini, Matteo and Pucci, Andrea and Ciaramella, Ernesto and Catani, Jacopo},
  title   = {Comparative Test of Novel Fluorescent Optical Antennas for {LED}- and Laser-Based Optical Wireless Communications},
  journal = {Advanced Optical Materials},
  volume  = {13},
  number  = {13},
  pages   = {2402367},
  year    = {2025},
  doi     = {10.1002/adom.202402367}
}

@misc{liu2026gnr,
  author       = {Liu, Xiaochen and Linnartz, Jean-Paul M. G. and Cunha, Thiago E. B.},
  title        = {Modeling and Modulation Optimization for {OWC} Limited by Electronic and Photonic Bandwidth},
  year         = {2026},
  eprint       = {2605.03976},
  archivePrefix = {arXiv},
  primaryClass = {eess.SP},
  howpublished = {arXiv:2605.03976},
  doi          = {10.48550/arXiv.2605.03976}
}

@misc{wang2026spad,
  author       = {Wang, Chen and Xu, Zhiyong and Wang, Jingyuan and Li, Jianhua and Mou, Weifeng and Zhu, Huatao},
  title        = {Unified Analytical Framework for {SPAD} Array Receivers with Dead-Time-Induced Blocking Loss and Inter-Symbol Interference in {PAM}-{OWC} Systems},
  year         = {2026},
  eprint       = {2605.28560},
  archivePrefix = {arXiv},
  primaryClass = {eess.SP},
  howpublished = {arXiv:2605.28560},
  doi          = {10.48550/arXiv.2605.28560}
}

@article{mardanikorani2020pam,
  author  = {Mardanikorani, Shokoufeh and Deng, Xiong and Linnartz, Jean-Paul M. G.},
  title   = {Optimization and Comparison of {M}-{PAM} and Optical {OFDM} Modulation for Optical Wireless Communication},
  journal = {IEEE Open Journal of the Communications Society},
  volume  = {1},
  pages   = {1721--1737},
  year    = {2020},
  doi     = {10.1109/OJCOMS.2020.3034204}
}

@article{barros2012ofdmpam,
  author  = {Barros, Daniel J. F. and Wilson, Sarah K. and Kahn, Joseph M.},
  title   = {Comparison of Orthogonal Frequency-Division Multiplexing and Pulse-Amplitude Modulation in Indoor Optical Wireless Links},
  journal = {IEEE Transactions on Communications},
  volume  = {60},
  number  = {1},
  pages   = {153--163},
  year    = {2012},
  doi     = {10.1109/TCOMM.2011.112311.100538}
}

@article{chaaban2022capacity,
  author  = {Chaaban, Anas and Rezki, Zouheir and Alouini, Mohamed-Slim},
  title   = {On the Capacity of Intensity-Modulation Direct-Detection {G}aussian Optical Wireless Communication Channels: A Tutorial},
  journal = {IEEE Communications Surveys {\&} Tutorials},
  volume  = {24},
  number  = {1},
  pages   = {455--491},
  year    = {2022},
  doi     = {10.1109/COMST.2021.3120087}
}

@article{deng2018lednl,
  author  = {Deng, Xiong and Mardanikorani, Shokoufeh and Wu, Yan and Arulandu, Kumar and Chen, Bin and Khalid, Amir M. and Linnartz, Jean-Paul M. G.},
  title   = {Mitigating {LED} Nonlinearity to Enhance Visible Light Communications},
  journal = {IEEE Transactions on Communications},
  volume  = {66},
  number  = {11},
  pages   = {5593--5607},
  year    = {2018},
  doi     = {10.1109/TCOMM.2018.2858239}
}

@misc{villenas2025rin,
  author       = {Villenas, Felipe and Wu, Kaiquan and G{\"u}ltekin, Yunus Can and Riani, Jamal and Alvarado, Alex},
  title        = {Beyond 200 {G}b/s/lane: An Analytical Approach to Optimal Detection in Shaped {IM}-{DD} Optical Links with Relative Intensity Noise},
  year         = {2025},
  eprint       = {2506.19684},
  archivePrefix = {arXiv},
  primaryClass = {eess.SP},
  howpublished = {arXiv:2506.19684},
  doi          = {10.48550/arXiv.2506.19684}
}

@article{wiegart2021ps4pam,
  author  = {Wiegart, Thomas and Da Ros, Francesco and Yankov, Metodi Plamenov and Steiner, Fabian and Gaiarin, Simone and Wesel, Richard D.},
  title   = {Probabilistically Shaped 4-{PAM} for Short-Reach {IM}/{DD} Links With a Peak Power Constraint},
  journal = {Journal of Lightwave Technology},
  volume  = {39},
  number  = {2},
  pages   = {400--405},
  year    = {2021},
  doi     = {10.1109/JLT.2020.3029371}
}

@article{gutema2022ps,
  author  = {Gutema, Tilahun Zerihun and Haas, Harald and Popoola, Wasiu O.},
  title   = {{WDM} Based 10.8 {G}bps Visible Light Communication With Probabilistic Shaping},
  journal = {Journal of Lightwave Technology},
  volume  = {40},
  number  = {15},
  pages   = {5062--5069},
  year    = {2022},
  doi     = {10.1109/JLT.2022.3175575}
}

@article{kafizov2024pcs,
  author  = {Kafizov, Amanat and Elzanaty, Ahmed and Alouini, Mohamed-Slim},
  title   = {Probabilistic Constellation Shaping for Enhancing Spectral Efficiency in {NOMA} {VLC} Systems},
  journal = {IEEE Transactions on Wireless Communications},
  volume  = {23},
  number  = {8},
  pages   = {9958--9971},
  year    = {2024},
  doi     = {10.1109/TWC.2024.3367442}
}

@article{shi2023pam4,
  author  = {Shi, Jianyang and Wei, Yuan and Luo, Zhiteng and Li, Ziwei and Shen, Chao and Zhang, Junwen and Chi, Nan},
  title   = {8.8 {G}bps {PAM}-4 visible light communication link using an external modulator and a neural network equalizer},
  journal = {Optics Letters},
  volume  = {48},
  number  = {20},
  pages   = {5193--5196},
  year    = {2023},
  doi     = {10.1364/OL.503822}
}

@article{alvarado2016fec,
  author  = {Alvarado, Alex and Agrell, Erik and Lavery, Domanic and Maher, Robert and Bayvel, Polina},
  title   = {Replacing the Soft-Decision {FEC} Limit Paradigm in the Design of Optical Communication Systems},
  journal = {Journal of Lightwave Technology},
  volume  = {34},
  number  = {2},
  pages   = {707--721},
  year    = {2016},
  doi     = {10.1109/JLT.2015.2482718}
}

@article{agrell2021recipes,
  author  = {Agrell, Erik and Secondini, Marco and Alvarado, Alex and Yoshida, Tsuyoshi},
  title   = {Performance Prediction Recipes for Optical Links},
  journal = {IEEE Photonics Technology Letters},
  volume  = {33},
  number  = {18},
  pages   = {1034--1037},
  year    = {2021},
  doi     = {10.1109/LPT.2021.3093790}
}

@article{chang2010fec,
  author  = {Chang, Frank and Onohara, Kiyoshi and Mizuochi, Takashi},
  title   = {Forward error correction for 100 {G} transport networks},
  journal = {IEEE Communications Magazine},
  volume  = {48},
  number  = {3},
  pages   = {S48--S55},
  year    = {2010},
  doi     = {10.1109/MCOM.2010.5434378}
}

@article{arfaoui2021channel,
  author  = {Arfaoui, Mohamed Amine and Soltani, Mohammad Dehghani and Tavakkolnia, Iman and Ghrayeb, Ali and Assi, Chadi M. and Safari, Majid and Haas, Harald},
  title   = {Measurements-Based Channel Models for Indoor {L}i{F}i Systems},
  journal = {IEEE Transactions on Wireless Communications},
  volume  = {20},
  number  = {2},
  pages   = {827--842},
  year    = {2021},
  doi     = {10.1109/TWC.2020.3028456}
}

@article{zhu2022gbsm,
  author  = {Zhu, Xiuming and Wang, Cheng-Xiang and Huang, Jie and Chen, Ming and Haas, Harald},
  title   = {A Novel 3{D} Non-Stationary Channel Model for 6{G} Indoor Visible Light Communication Systems},
  journal = {IEEE Transactions on Wireless Communications},
  volume  = {21},
  number  = {10},
  pages   = {8292--8307},
  year    = {2022},
  doi     = {10.1109/TWC.2022.3165569}
}

@article{miramirkhani2023bb,
  author  = {Miramirkhani, Farshad and Baykas, Tuncer and Elamassie, Mohammed and Uysal, Murat},
  title   = {{IEEE} 802.11bb Reference Channel Models for Light Communications},
  journal = {IEEE Communications Standards Magazine},
  volume  = {7},
  number  = {4},
  pages   = {84--89},
  year    = {2023},
  doi     = {10.1109/MCOMSTD.0006.2300009}
}

@article{ma2024mobile,
  author  = {Ma, Shuai and Sheng, Haihong and Sun, Junchang and Li, Hang and Liu, Xiaodong and Qiu, Chen and Safari, Majid and Al-Dhahir, Naofal and Li, Shiyin},
  title   = {Feasibility Conditions for Mobile {L}i{F}i},
  journal = {IEEE Transactions on Wireless Communications},
  volume  = {23},
  number  = {7},
  pages   = {7911--7923},
  year    = {2024},
  doi     = {10.1109/TWC.2023.3346056}
}

@article{miramirkhani2020channel,
  author  = {Miramirkhani, Farshad and Uysal, Murat},
  title   = {Channel modelling for indoor visible light communications},
  journal = {Philosophical Transactions of the Royal Society A: Mathematical, Physical and Engineering Sciences},
  volume  = {378},
  number  = {2169},
  pages   = {20190187},
  year    = {2020},
  doi     = {10.1098/rsta.2019.0187}
}

@article{soltani2019orientation,
  author  = {Soltani, Mohammad Dehghani and Purwita, Ardimas Andi and Zeng, Zhihong and Haas, Harald and Safari, Majid},
  title   = {Modeling the Random Orientation of Mobile Devices: Measurement, Analysis and {L}i{F}i Use Case},
  journal = {IEEE Transactions on Communications},
  volume  = {67},
  number  = {3},
  pages   = {2157--2172},
  year    = {2019},
  doi     = {10.1109/TCOMM.2018.2882213}
}

@article{chen2021uplink,
  author  = {Chen, Cheng and Basnayaka, Dushyantha A. and Purwita, Ardimas Andi and Wu, Xiping and Haas, Harald},
  title   = {Wireless Infrared-Based {L}i{F}i Uplink Transmission With Link Blockage and Random Device Orientation},
  journal = {IEEE Transactions on Communications},
  volume  = {69},
  number  = {2},
  pages   = {1175--1188},
  year    = {2021},
  doi     = {10.1109/TCOMM.2020.3035405}
}

@inproceedings{mana2021multilink,
  author    = {Maravanchery Mana, Sreelal and Gabra, Kerolos Gabra Kamel and Mohammadi Kouhini, Sepideh and Hellwig, Peter and Hilt, Jonas and Jungnickel, Volker},
  title     = {An Efficient Multi-Link Channel Model for {L}i{F}i},
  booktitle = {2021 IEEE 32nd Annual International Symposium on Personal, Indoor and Mobile Radio Communications (PIMRC)},
  pages     = {1--6},
  year      = {2021},
  doi       = {10.1109/PIMRC50174.2021.9569661}
}

@article{tong2023industrial,
  author  = {Tong, Yu and Tang, Pan and Zhang, Jianhua and Liu, Shuo and Yin, Yue and Liu, Baoling and Xia, Liang},
  title   = {Channel Characteristics and Link Adaption for Visible Light Communication in an Industrial Scenario},
  journal = {Sensors},
  volume  = {23},
  number  = {7},
  pages   = {3442},
  year    = {2023},
  doi     = {10.3390/s23073442}
}

@article{celik2023survey,
  author  = {Celik, Abdulkadir and Romdhane, Imene and Kaddoum, Georges and Eltawil, Ahmed M.},
  title   = {A Top-Down Survey on Optical Wireless Communications for the Internet of Things},
  journal = {IEEE Communications Surveys {\&} Tutorials},
  volume  = {25},
  number  = {1},
  pages   = {1--45},
  year    = {2023},
  doi     = {10.1109/COMST.2022.3220504}
}

@article{khorov2022bb,
  author  = {Khorov, Evgeny and Levitsky, Ilya},
  title   = {Current Status and Challenges of {L}i-{F}i: {IEEE} 802.11bb},
  journal = {IEEE Communications Standards Magazine},
  volume  = {6},
  number  = {2},
  pages   = {35--41},
  year    = {2022},
  doi     = {10.1109/MCOMSTD.0001.2100104}
}

@article{wu2021hybrid,
  author  = {Wu, Xiping and Soltani, Mohammad Dehghani and Zhou, Lai and Safari, Majid and Haas, Harald},
  title   = {Hybrid {L}i{F}i and {W}i{F}i Networks: A Survey},
  journal = {IEEE Communications Surveys {\&} Tutorials},
  volume  = {23},
  number  = {2},
  pages   = {1398--1420},
  year    = {2021},
  doi     = {10.1109/COMST.2021.3058296}
}

@article{soltani2023lifi2,
  author  = {Soltani, Mohammad Dehghani and Qidan, Ahmad Adnan and Huang, Sicong and Yosuf, Barzan and Mohamed, Sanaa and Singh, Ravinder and Liu, Yizhe and Ali, Wajahat and Chen, Rui and Kazemi, Hossein and Sarbazi, Elham and Berde, Bruno and Chiaroni, Dominique and Bechadergue, Bastien and Abdel-Dayem, Fatma and Soni, Harsh and Tabu, Jose and Perrufel, Micheline and Serafimovski, Nikola and El-Gorashi, Taisir E. H. and Elmirghani, Jaafar and Crisp, Michael and Penty, Richard and White, Ian H. and Haas, Harald and Safari, Majid},
  title   = {Terabit Indoor Laser-Based Wireless Communications: {L}i{F}i 2.0 for 6{G}},
  journal = {IEEE Wireless Communications},
  volume  = {30},
  number  = {5},
  pages   = {36--43},
  year    = {2023},
  doi     = {10.1109/MWC.007.2300121}
}

@article{bober2024ojvt,
  author  = {Bober, Kai Lennert and Ebmeyer, Anselm and Dressler, Falko and Freund, Ronald and Jungnickel, Volker},
  title   = {{L}i{F}i for Industry 4.0: Main Features, Implementation and Initial Testing of {IEEE} {S}td 802.15.13},
  journal = {IEEE Open Journal of Vehicular Technology},
  volume  = {5},
  pages   = {1625--1636},
  year    = {2024},
  doi     = {10.1109/OJVT.2024.3481884}
}

@article{ji2025usercentric,
  author  = {Ji, Han and Wu, Xiping},
  title   = {Resource and Mobility Management in Hybrid {L}i{F}i and {W}i{F}i Networks: A User-Centric Learning Approach},
  journal = {IEEE Transactions on Wireless Communications},
  volume  = {24},
  number  = {2},
  pages   = {1293--1305},
  year    = {2025},
  doi     = {10.1109/TWC.2024.3507828}
}

@article{ji2024atcnn,
  author  = {Ji, Han and Wu, Xiping and Wang, Qiang and Redmond, Stephen J. and Tavakkolnia, Iman},
  title   = {Adaptive Target-Condition Neural Network: {DNN}-Aided Load Balancing for Hybrid {L}i{F}i and {W}i{F}i Networks},
  journal = {IEEE Transactions on Wireless Communications},
  volume  = {23},
  number  = {7},
  pages   = {7307--7318},
  year    = {2024},
  doi     = {10.1109/TWC.2023.3339503}
}

@article{ullah2021ns3,
  author  = {Ullah, Shakir and Rehman, Saeed Ur and Chong, Peter Han Joo},
  title   = {A Comprehensive Open-Source Simulation Framework for {L}i{F}i Communication},
  journal = {Sensors},
  volume  = {21},
  number  = {7},
  pages   = {2485},
  year    = {2021},
  doi     = {10.3390/s21072485}
}

@article{aldalbahi2017ns3,
  author  = {Aldalbahi, Adel and Rahaim, Michael and Khreishah, Abdallah and Ayyash, Moussa and Little, Thomas D. C.},
  title   = {Visible Light Communication Module: An Open Source Extension to the ns3 Network Simulator With Real System Validation},
  journal = {IEEE Access},
  volume  = {5},
  pages   = {22144--22158},
  year    = {2017},
  doi     = {10.1109/ACCESS.2017.2759779}
}

@article{haas2020indoor,
  author  = {Haas, Harald and Yin, Liang and Chen, Cheng and Videv, Stefan and Parol, Damian and Poves, Enrique and Alshaer, Hamada and Islim, Mohamed Sufyan},
  title   = {Introduction to indoor networking concepts and challenges in {L}i{F}i},
  journal = {IEEE/OSA Journal of Optical Communications and Networking},
  volume  = {12},
  number  = {2},
  pages   = {A190--A203},
  year    = {2020},
  doi     = {10.1364/JOCN.12.00A190}
}

@article{murad2022handover,
  author  = {Murad, Sallar Salam and Yussof, Salman and Hashim, Wahidah and Badeel, Rozin},
  title   = {Three-Phase Handover Management and Access Point Transition Scheme for Dynamic Load Balancing in Hybrid {L}i{F}i/{W}i{F}i Networks},
  journal = {Sensors},
  volume  = {22},
  number  = {19},
  pages   = {7583},
  year    = {2022},
  doi     = {10.3390/s22197583}
}

@article{linnartz2022eliot,
  author  = {Linnartz, Jean-Paul M. G. and Corr{\^e}a, Camilo R. B. and Cunha, Thiago E. B. and Tangdiongga, Eduward and Koonen, Ton and Deng, Xiong and Wendt, Matthias and Abbo, Anteneh A. and Stobbelaar, Peter J. and Polak, Petr and M{\"u}ller, Marcus and Behnke, Dennis and Mart{\'i}nez, Manuel and Vicent, Salvador and Metin, Tolga and Emmelmann, Marc and Kouhini, Sepideh Mohammadi and Bober, Kai Lennert and Kottke, Christoph and Jungnickel, Volker},
  title   = {{ELI}o{T}: enhancing {L}i{F}i for next-generation {I}nternet of things},
  journal = {EURASIP Journal on Wireless Communications and Networking},
  volume  = {2022},
  number  = {1},
  pages   = {89},
  year    = {2022},
  doi     = {10.1186/s13638-022-02168-6}
}

@misc{ancillotti2025cots,
  author       = {Ancillotti, Emilio and Pescosolido, Loreto and Passarella, Andrea},
  title        = {Toward Hybrid {COTS}-based {L}i{F}i/{W}i{F}i Networks with {Q}o{S} Requirements in Mobile Environments},
  year         = {2025},
  eprint       = {2511.00210},
  archivePrefix = {arXiv},
  primaryClass = {cs.NI},
  howpublished = {arXiv:2511.00210},
  doi          = {10.48550/arXiv.2511.00210}
}

@article{wheatley2026iec,
  author  = {Wheatley, Trevor and Meng, Ying and Wang, Yunfeng and Bi, Wanjun and Gutoehrlein, Katja and Frederiksen, Annette},
  title   = {Challenges, development, and use of the laser safety standard {IEC} 60825-1:2014},
  journal = {Journal of Laser Applications},
  volume  = {38},
  number  = {1},
  pages   = {012043},
  year    = {2026},
  doi     = {10.2351/7.0001804}
}

@inproceedings{schulmeister2015extended,
  author    = {Schulmeister, Karl},
  title     = {Classification of extended source products according to {IEC} 60825-1},
  booktitle = {ILSC 2015: Proceedings of the International Laser Safety Conference},
  publisher = {Laser Institute of America},
  pages     = {271--280},
  year      = {2015},
  doi       = {10.2351/1.5056849}
}

@article{kumar2024speckle,
  author  = {Kumar, Virendra and Sharma, Parag and Mehta, Dalip Singh},
  title   = {Design and development of speckle-free high-power laser-driven phosphor converted compact automotive headlamp module},
  journal = {Journal of Physics: Photonics},
  volume  = {6},
  number  = {2},
  pages   = {025008},
  year    = {2024},
  doi     = {10.1088/2515-7647/ad2bd2}
}

@article{jensen2025speckle,
  author  = {Jensen, Ole Bjarlin and Xu, Jian and Jakobsen, Michael Linde},
  title   = {Dependence of speckle contrast in laser lighting on laser and phosphor characteristics},
  journal = {Journal of Luminescence},
  volume  = {286},
  pages   = {121415},
  year    = {2025},
  doi     = {10.1016/j.jlumin.2025.121415}
}

@article{rahman2022laserphosphor,
  author  = {Rahman, Faiz},
  title   = {Diode laser-excited phosphor-converted light sources: a review},
  journal = {Optical Engineering},
  volume  = {61},
  number  = {6},
  pages   = {060901},
  year    = {2022},
  doi     = {10.1117/1.OE.61.6.060901}
}

@article{congar2018rin,
  author  = {Congar, Antoine and Hussain, Kamal and Pareige, Christelle and Butt{\'e}, Rapha{\"e}l and Grandjean, Nicolas and Besnard, Pascal and Trebaol, St{\'e}phane},
  title   = {Impact of Mode-Hopping Noise on {I}n{G}a{N} Edge Emitting Laser Relative Intensity Noise Properties},
  journal = {IEEE Journal of Quantum Electronics},
  volume  = {54},
  number  = {1},
  pages   = {1100107},
  year    = {2018},
  doi     = {10.1109/JQE.2017.2774358}
}

@article{elfar2026vlcnoise,
  author  = {ElFar, Sara H. and Ikki, Salama and Sebak, Abdel Razik},
  title   = {Performance Analysis of {VLC} Systems Over Random Channels With Signal-Dependent Shot Noise and Relative Intensity Noise},
  journal = {IEEE Photonics Journal},
  volume  = {18},
  number  = {4},
  pages   = {7301615},
  year    = {2026},
  doi     = {10.1109/JPHOT.2026.3714407}
}

@article{chi2015diffuser,
  author  = {Chi, Yu-Chieh and Hsieh, Dan-Hua and Lin, Chung-Yu and Chen, Hsiang-Yu and Huang, Chia-Yen and He, Jr-Hau and Ooi, Boon and DenBaars, Steven P. and Nakamura, Shuji and Kuo, Hao-Chung and Lin, Gong-Ru},
  title   = {Phosphorous Diffuser Diverged Blue Laser Diode for Indoor Lighting and Communication},
  journal = {Scientific Reports},
  volume  = {5},
  pages   = {18690},
  year    = {2015},
  doi     = {10.1038/srep18690}
}

@article{ata2026survey,
  author  = {Ata, Yalcin and Al-Sallami, Farah Mahdi and Gokce, Muhsin Caner and Vegni, Anna Maria and Rajbhandari, Sujan and Baykal, Yahya},
  title   = {Optical Wireless Communication in Atmosphere and Underwater: Statistical Models, Improvement Techniques, and Recent Applications},
  journal = {IEEE Communications Surveys {\&} Tutorials},
  volume  = {28},
  pages   = {4248--4284},
  year    = {2026},
  doi     = {10.1109/COMST.2025.3649735}
}

@article{chow2024jlt6g,
  author  = {Chow, Chi-Wai},
  title   = {Recent Advances and Future Perspectives in Optical Wireless Communication, Free Space Optical Communication and Sensing for 6{G}},
  journal = {Journal of Lightwave Technology},
  volume  = {42},
  number  = {11},
  pages   = {3972--3980},
  year    = {2024},
  doi     = {10.1109/JLT.2024.3386630}
}

@article{krishnamoorthy2025owc,
  author  = {Krishnamoorthy, Anil and Safi, Hossein and Younus, Othman and Kazemi, Hossein and Osahon, Isaac N. O. and Liu, Mingqing and Liu, Yizhe and Babadi, Sanaz and Ahmad, Rizwan and Ihsan, Asim and Majlesein, Beatriz and Huang, Yi and Herrnsdorf, Johannes and Rajbhandari, Sujan and McKendry, Jonathan J. D. and Tavakkolnia, Iman and Caglayan, Humeyra and Helmers, Henning and Turnbull, Graham and Samuel, Ifor D. W. and Dawson, Martin D. and Schober, Robert and Haas, Harald},
  title   = {Optical Wireless Communications: Enabling the Next-Generation Network of Networks},
  journal = {IEEE Vehicular Technology Magazine},
  volume  = {20},
  number  = {2},
  pages   = {20--39},
  year    = {2025},
  doi     = {10.1109/MVT.2025.3555366}
}

@article{lu2025underwater,
  author  = {Lu, Zhilan and Li, Zhenhao and Lin, Xianhao and Cai, Jifan and Li, Fujie and Xu, Zengyi and Wang, Lai and Zhou, Yingjun and Shen, Chao and Zhang, Junwen and Chi, Nan},
  title   = {170 {G}bps {PDM} underwater visible light communication utilizing a compact 5-{$\lambda$} laser transmitter and a reciprocal differential receiver},
  journal = {Photonics Research},
  volume  = {13},
  number  = {6},
  pages   = {1654},
  year    = {2025},
  doi     = {10.1364/PRJ.551924}
}

@misc{ahmad2026arxiv,
  author       = {Ahmad, Hussain and Gillani, Syed Muhammad Talha and Omer, Toheed and Aslam, Saleem},
  title        = {Laser-Diode {LiFi} With Diffused-Beam Optics: System-Level Modeling and a Cross-Validated ns-3 Simulation Framework},
  howpublished = {arXiv:2608.10950 [eess.SP]},
  year         = {2026},
  note         = {Earlier version of the present work; superseded by this article}
}

@article{sackinger2010limit,
  author  = {Eduard S{\"a}ckinger},
  title   = {The Transimpedance Limit},
  journal = {IEEE Transactions on Circuits and Systems I: Regular Papers},
  volume  = {57},
  number  = {8},
  pages   = {1848-1856},
  year    = {2010},
  doi     = {10.1109/TCSI.2009.2037847}
}

@book{sackinger2017tia,
  author  = {Eduard S{\"a}ckinger},
  title   = {Analysis and Design of Transimpedance Amplifiers for Optical Receivers},
  publisher = {Wiley},
  year    = {2017},
  doi     = {10.1002/9781119264422}
}

@article{mohan2000bw,
  author  = {Sunderarajan S. Mohan and Maria del Mar Hershenson and Stephen P. Boyd and Thomas H. Lee},
  title   = {Bandwidth Extension in {CMOS} with Optimized On-Chip Inductors},
  journal = {IEEE Journal of Solid-State Circuits},
  volume  = {35},
  number  = {3},
  pages   = {346-355},
  year    = {2000},
  doi     = {10.1109/4.826816}
}

@article{personick1973rx,
  author  = {Stewart D. Personick},
  title   = {Receiver Design for Digital Fiber Optic Communication Systems, {I}},
  journal = {Bell System Technical Journal},
  volume  = {52},
  number  = {6},
  pages   = {843-874},
  year    = {1973},
  doi     = {10.1002/j.1538-7305.1973.tb01993.x}
}

@article{muoi1984rx,
  author  = {Tran Van Muoi},
  title   = {Receiver Design for High-Speed Optical-Fiber Systems},
  journal = {Journal of Lightwave Technology},
  volume  = {2},
  number  = {3},
  pages   = {243-267},
  year    = {1984},
  doi     = {10.1109/JLT.1984.1073617}
}

@article{hullett1976tia,
  author  = {J. L. Hullett and T. V. Muoi},
  title   = {A Feedback Receive Amplifier for Optical Transmission Systems},
  journal = {IEEE Transactions on Communications},
  volume  = {24},
  number  = {10},
  pages   = {1180-1185},
  year    = {1976},
  doi     = {10.1109/TCOM.1976.1093224}
}

@inproceedings{kassem2019tia,
  author  = {Amany Kassem and Izzat Darwazeh},
  title   = {A High Bandwidth Modified Regulated Cascode {TIA} for High Capacitance Photodiodes in {VLC}},
  booktitle = {Proc. IEEE International Symposium on Circuits and Systems (ISCAS), Sapporo, Japan},
  year    = {2019},
  doi     = {10.1109/ISCAS.2019.8702692}
}

@inproceedings{cura2013tia,
  author  = {Jos{\'e} Luis Cura and Luis Nero Alves},
  title   = {Bandwidth Improvements in Transimpedance Amplifiers for Visible-Light Receiver Front-Ends},
  booktitle = {Proc. 2013 IEEE 20th International Conference on Electronics, Circuits, and Systems (ICECS)},
  pages   = {831-834},
  year    = {2013},
  doi     = {10.1109/ICECS.2013.6815543}
}

@misc{iec62471_2006,
  author       = {{International Electrotechnical Commission}},
  title        = {Photobiological Safety of Lamps and Lamp Systems},
  howpublished = {IEC 62471:2006 (CIE S 009:2002), Ed. 1.0, Geneva, Switzerland},
  year         = {2006}
}

@misc{iec62471_7,
  author       = {{International Electrotechnical Commission}},
  title        = {Photobiological Safety of Lamps and Lamp Systems---Part 7: Light Sources and Luminaires Primarily Emitting Visible Radiation},
  howpublished = {IEC 62471-7:2023, Ed. 1.0, Geneva, Switzerland},
  year         = {2023}
}

@inproceedings{schulmeister2017illum,
  author  = {Karl Schulmeister and Jan Daem},
  title   = {Classification of laser illuminated light sources under {IEC} 60825-1 edition 3},
  booktitle = {Proc. ILSC 2017: Proceedings of the International Laser Safety Conference, Laser Institute of America},
  pages   = {174-180},
  year    = {2017},
  doi     = {10.2351/1.5056873}
}

@article{schulmeister2016rg2,
  author  = {Karl Schulmeister and Jan Daem},
  title   = {Risk of retinal injury from "Risk Group 2" laser illuminated projectors},
  journal = {Journal of Laser Applications},
  volume  = {28},
  number  = {4},
  pages   = {042002},
  year    = {2016},
  doi     = {10.2351/1.4954930}
}

@article{wu2017eyesafe,
  author  = {Tsai-Chen Wu and Yu-Chieh Chi and Huai-Yung Wang and Cheng-Ting Tsai and Yu-Fang Huang and Gong-Ru Lin},
  title   = {Tricolor {R/G/B} Laser Diode Based Eye-Safe White Lighting Communication Beyond 8 Gbit/s},
  journal = {Scientific Reports},
  volume  = {7},
  number  = {11},
  pages   = {11},
  year    = {2017},
  doi     = {10.1038/s41598-017-00052-8}
}

@book{oberkampf2010vv,
  author  = {William L. Oberkampf and Christopher J. Roy},
  title   = {Verification and Validation in Scientific Computing},
  publisher = {Cambridge University Press, Cambridge, UK},
  year    = {2010},
  doi     = {10.1017/CBO9780511760396}
}

@misc{aiaa1998vv,
  author       = {{American Institute of Aeronautics and Astronautics}},
  title        = {Guide for the Verification and Validation of Computational Fluid Dynamics Simulations},
  howpublished = {AIAA Standard G-077-1998, Reston, VA},
  year         = {1998}
}

@article{sargent2013vv,
  author  = {Robert G. Sargent},
  title   = {Verification and validation of simulation models},
  journal = {Journal of Simulation},
  volume  = {7},
  number  = {1},
  pages   = {12-24},
  year    = {2013},
  doi     = {10.1057/jos.2012.20}
}

@misc{asme2019vv,
  author       = {{American Society of Mechanical Engineers}},
  title        = {Standard for Verification and Validation in Computational Solid Mechanics},
  howpublished = {ASME V\&V 10-2019 (R2025), New York, NY},
  year         = {2019}
}

@article{roache1998vv,
  author  = {Patrick J. Roache},
  title   = {Verification of Codes and Calculations},
  journal = {AIAA Journal},
  volume  = {36},
  number  = {5},
  pages   = {696-702},
  year    = {1998},
  doi     = {10.2514/2.457}
}

@article{maraqa2024starris,
  author  = {Omar Maraqa and Sylvester Aboagye and Telex M. N. Ngatched},
  title   = {Optical {STAR-RIS}-Aided {VLC} Systems: {RSMA} Versus {NOMA}},
  journal = {IEEE Open Journal of the Communications Society},
  volume  = {5},
  pages   = {430--441},
  year    = {2024},
  doi     = {10.1109/OJCOMS.2023.3347534}
}

@article{surampudi2018attocell,
  author  = {Atchutananda Surampudi and Radha Krishna Ganti},
  title   = {Interference Characterization in Downlink {Li-Fi} Optical Attocell Networks},
  journal = {Journal of Lightwave Technology},
  volume  = {36},
  number  = {16},
  pages   = {3211--3228},
  year    = {2018},
  doi     = {10.1109/JLT.2018.2836932}
}

@article{tabassum2018coverage,
  author  = {Hina Tabassum and Ekram Hossain},
  title   = {Coverage and Rate Analysis for Co-Existing {RF}/{VLC} Downlink Cellular Networks},
  journal = {IEEE Transactions on Wireless Communications},
  volume  = {17},
  number  = {4},
  pages   = {2588--2601},
  year    = {2018},
  doi     = {10.1109/TWC.2018.2799204}
}

@inproceedings{mach2017d2d,
  author  = {Pavel Mach and Zdenek Becvar and Mehyar Najla and Stanislav Zvanovec},
  title   = {Combination of visible light and radio frequency bands for device-to-device communication},
  booktitle = {Proc. Proc. IEEE 28th Annual International Symposium on Personal, Indoor and Mobile Radio Communications (PIMRC)},
  pages   = {1--7},
  year    = {2017},
  doi     = {10.1109/PIMRC.2017.8292746}
}

\end{document}